\documentclass[aps,preprint,nofootinbib,hyperref]{revtex4}%
\usepackage{amsfonts}
\usepackage{amsmath}
\usepackage{amssymb}
\usepackage{hyperref}
\usepackage{graphicx}%
\providecommand{\U}[1]{\protect\rule{.1in}{.1in}}
\providecommand{\U}[1]{\protect\rule{.1in}{.1in}}
\providecommand{\U}[1]{\protect\rule{.1in}{.1in}}
\providecommand{\U}[1]{\protect\rule{.1in}{.1in}}
\providecommand{\U}[1]{\protect\rule{.1in}{.1in}}
\providecommand{\U}[1]{\protect\rule{.1in}{.1in}}
\providecommand{\U}[1]{\protect\rule{.1in}{.1in}}

\begin{document}
\preprint{ }
\title{Wavefunction for the Universe Circa the Beginning \\with Dynamically Determined Unique Initial Conditions}
\author{Itzhak Bars}
\affiliation{Department of Physics and Astronomy, University of Southern California, Los
Angeles, CA, 90089-0484, USA,}
\affiliation{Perimeter Institute for Theoretical Physics, Waterloo, ON, N2L 2Y5, Canada}

\begin{abstract}
In this paper I will first outline an effective field theory for cosmology
(EFTC) that is based on the Standard Model coupled to General Relativity and
improved with Weyl symmetry. There are no new physical degrees of freedom in
this theory, but what is new is an enlargement of the domain of the existing
physical fields and of spacetime via the larger symmetry, thus curing the
geodesic incompleteness of the traditional theory. Invoking the softer
behavior of an underlying theory of quantum gravity, I further argue that it
is reasonable to ban higher curvature terms in the effective action, thus
making this EFTC mathematically well behaved at gravitational singularities,
as well as geodesically complete, thus able to make new physics predictions.
Using this EFTC, I show some predictions of surprising behavior of the
universe at singularities including a unique set of big-bang initial
conditions that emerge from a dynamical attractor mechanism. I will illustrate
this behavior with detailed formulas and plots of the classical solutions and
the quantum wavefunction that are continuous across singularities for a
cosmology that includes the past and future of the big bang. The solutions are
given in the geodesically complete global mini-superspace that is similar to
the extended spacetime of a black hole or extended Rindler spacetime. The
analytic continuation of the quantum wavefunction across the horizons
describes the passage through the singularities. This analytic continuation
solves a long-standing problem of the singular $(-1/r^{2})$ potential in
quantum mechanics that dates back to Von Neumann. The analytic properties of
the wavefunction also reveal an infinite stack of universes sewn together at
the horizons of the geodesically complete space. Finally a comparison with
recent papers using the path integral approach in cosmology is given.

\end{abstract}

\maketitle
\tableofcontents

\newpage

\section{Introduction}

This paper presents an extension of work I started ten years ago in the
context of 2T-Physics \cite{barsGrav2T}\cite{BC}, and pursued in a series of
papers on cosmology and black holes in collaboration with Chen, Steinhardt,
Turok, Araya and James, where the role of Weyl symmetry, in the geodesically
complete form that emerges from 2T-physics, was emphasized \cite{BC}%
-\cite{barsArayaBH}. By now, foundational ideas are better understood and in
this paper applied to the quantum wavefunction for the universe. The current
paper highlights the main concepts and new results on classical cosmological
solutions, the quantum wavefunction and associated propagator.

The paper is organized as follows. Section \ref{theory} introduces the
geodesically complete fundamental theory and its attractive features, while
section \ref{superspace} discusses its mini-superspace, its geometrical
structure and the transformation between systems of mini-superspace
coordinates that highlights a global system analogous to the Kruskal-Szekeres
global coordinates for a black hole. In section \ref{attractor} explicit
analytic classical solutions of the mini-superspace are given; these display
an attractor mechanism leading to unique dynamically determined initial
conditions at the big bang, and help establish a theorem on the behavior of
all the degrees of freedom at cosmological singularities. The Wheeler de Witt
equation (WdWe) that also leads to the same attractor mechanism is solved
analytically in three stages. First, in section \ref{QM1} the WdWe is setup
using geodesically complete global coordinates, quantum ordering is settled
globally, a 2-step approximation scheme is devised, and the general physical
behavior of the wavefunction is qualitatively determined through an effective
potential in a Schr\"{o}dinger-like equation. Second, in section \ref{QM2} the
continuity of the wavefunction is determined across the horizons in the global
mini-superspace. Third, in section \ref{QM3} the full solution for the
wavefunction containing no unknown parameters is explicitly given, and its
predicted form at the big bang is displayed. Finally an overall discussion is
given in section \ref{outlook}; this highlights the results of this paper,
outlines areas for future progress, and contrasts this work to other recent
papers that discuss the quantization of mini-superspace in the path integral
approach, including the quantum wavefunction and propagators.

\section{The fundamental theory \label{theory}}

The Lagrangian for the standard model coupled to general relativity and
improved with Weyl symmetry to obtain the \textit{geodesically complete}
version of this theory, without adding new \textit{physical degrees of
freedom}, is \cite{BST}
\begin{equation}
\mathcal{L}\left(  x\right)  =\sqrt{-g}\left(
\begin{array}
[c]{c}%
L_{\text{SM}}\left(  A_{\mu}^{\gamma,W,Z,g},\;\psi_{q,l},\;\nu_{R}%
,\;\chi\right)  +g^{\mu\nu}\left(  \frac{1}{2}\partial_{\mu}\phi\partial_{\nu
}\phi-D_{\mu}H^{\dagger}D_{\nu}H\right) \\
-\left(  \frac{\lambda}{4}\left(  H^{\dagger}H-w^{2}\phi^{2}\right)
^{2}+\frac{\lambda^{\prime}}{4}\phi^{4}\right)  +\frac{1}{12}\left(  \phi
^{2}-2H^{\dagger}H\right)  R\left(  g\right)
\end{array}
\right)  . \label{action}%
\end{equation}
In the first line, $L_{\text{SM}}$ contains the usual degrees of freedom of
the extended standard model minimally coupled to gravity, namely, gauge bosons
$A_{\mu}^{\gamma,W,Z,g}$, quarks and leptons $\psi_{q,l}$, right handed
neutrinos $\nu_{R}$, some candidate(s) for dark matter $\chi,$ and their
SU$\left(  3\right)  \times$SU$\left(  2\right)  \times$U$\left(  1\right)  $
invariant interactions with the Higgs doublet $H,$ as well as the additional
singlet boson $\phi$ that can couple only to $\nu_{R},\chi$ because of the
electroweak gauge symmetry. The remaining terms in (\ref{action}) give the
kinetic terms for the conformally coupled scalars $\left(  \phi,H\right)  $,
their renormalizable and scale invariant potential energy capable of
dynamically generating the Higgs mass \cite{IB-higgsPortal}, and their locally
scale invariant unique non-minimal couplings to curvature $R\left(  g\right)
$. Under the local $\lambda\left(  x\right)  $ scale transformations,
$g_{\mu\nu}\rightarrow\lambda^{-2}g_{\mu\nu},~\phi\rightarrow\lambda
\phi,\;H\rightarrow\lambda H,$ $\psi_{q,l}\rightarrow\lambda^{3/2}\psi
_{q,l},\;A_{\mu}^{\gamma,W,Z,g}\rightarrow unchanged,$ the Lagrangian
(\ref{action}) transforms to a total derivative and therefore the action is
invariant. Because the Weyl symmetry can remove one gauge degree of freedom.
This version of the standard model coupled to gravity has \textit{no new
physical degrees of freedom although there is new physics because the domain
of the physical fields are considerably enlarged}. This action cannot contain
any dimensionful constants. Weyl invariant renormalization\footnote{In this
Weyl invariant renormalization scheme, the usual trace anomaly, of the energy
momentum tensor of all matter except $\phi,$ is still present, but it is
cancelled by an equal anomaly due to the additional term in the full energy
momentum tensor containing the extra field $\phi$ \cite{Percacci}. Thus, the
local Weyl symmetry survives in the quantized theory. \label{renorm}} of the
non-gravitational part of this action maintains the local Weyl symmetry by
taking the renormalization scale to be the field $\phi,$ thus allowing only
those counterterms that run as a function of Weyl invariant logarithms such as
$\ln\left(  H^{\dagger}H/\phi^{2}\right)  $ \cite{BST}\cite{BST-Higgs}.

Although there exist in the literature other forms of Weyl invariant field
couplings to gravity (in particular using the \textquotedblleft Stuckelberg
trick\textquotedblright), usually those are geodesically incomplete.
Incompleteness is a sign of unwittingly suppressing physical effects and
should be considered to be a serious problem looking for a cure. As discussed
in \cite{BC}-\cite{barsArayaBH}, in the case of only two scalar fields, the
form in (\ref{action}) is unique and geodesically complete, furthermore all
other incomplete forms can be obtained from this one by field redefinitions
\cite{BST} and artificially deleting patches of field space. In (\cite{BST})
it is shown how more scalar fields can be included in the geodesically
complete theory. In the current paper, I continue to explore the possibility
that the minimal case (\ref{action}) may be sufficient.

One of the virtues of this formalism is that it explains how the dimensionful
constants that fill the universe emerge from the same source. This is seen by
choosing a Weyl gauge, dubbed \textquotedblleft c-gauge\textquotedblright%
\ \cite{barsGrav2T}\cite{BST} that fixes $\phi\left(  x\right)  =\phi_{0}$ (a
constant) for all $x^{\mu}.$ Although several other gauge choices
\cite{nontrivialString} are convenient for various computations of
\textit{gauge invariants}, the c-gauge is most convenient to recognize the low
energy physics. In the c-gauge, the usual standard model with \textit{no
additional degrees of freedom}, containing all low energy dimensionful
parameters, is seen to arise from interactions with the scalars $\left(
\phi,H\right)  .$ In particular, the gravitational constant $G,$ the Higgs
vacuum value, and cosmological constant $\Lambda$, are%
\begin{equation}
\left(  16\pi G\right)  ^{-1}=\phi_{0}^{2}/12,\;\langle H^{\dagger}%
H\rangle=w^{2}\phi_{0}^{2},\;\left(  16\pi G\right)  ^{-1}2\Lambda
=\frac{\lambda^{\prime}}{4}\phi_{0}^{4}. \label{universal}%
\end{equation}
Universe-filling constants such as these raise the question whether these are
independent or related to each other. There is no literature that analyses
this question of cosmological significance. It is hard to imagine three
different mechanisms that would generate such an outcome. In the current
formalism, although the hierarchy of scales (which is achieved through
dimensionless parameters) is not explained, a unique source for all
universe-filling dimensionful parameters is identified. That such universal
parameters are not independent but are actually related to the same source,
resolves a long-standing puzzle for this author, thus providing more credence
to the current approach with Weyl symmetry.

Another significant feature introduced by the Weyl symmetry is the coefficient
of curvature, $\frac{1}{12}\left(  \phi^{2}-2H^{\dagger}H\right)  R\left(
g\right)  $, or a gauge fixed version such as the c-gauge, $\left(  \left(
16\pi G\right)  ^{-1}-\frac{2}{12}H^{\dagger}\left(  x\right)  H\left(
x\right)  \right)  R\left(  g\left(  x\right)  \right)  $. This relative sign
is obligatory and cannot be altered (otherwise a positive gravitational
constant is not possible) \cite{BST}. The question arises whether the dynamics
of the theory forces the sign to flip in some regions of spacetime $x^{\mu}$.
It was found through analytic solutions of the equations of motion that in
fact such a sign flip is the generic behavior \cite{BC}-\cite{barsArayaBH}. In
patches of spacetime where the sign is negative, gravity is repulsive, hence
antigravity rules in those regions of spacetime. The \textit{sign flip from
positive to negative can occur only at gravitational singularities} (see
explanation in Eq.(\ref{z})), therefore from the perspective of observers like
us in the gravity sector(s), antigravity occurs only on the other side of
cosmological or black hole type singularities. For geodesic completeness, all
gravity and antigravity patches must be included.

This is the structure predicted by the symmetries of 2T-physics for
relativistic 1T-physics \cite{barsGrav2T} and it was one of the main reasons
to start an investigation of this topic in 2008. It turns out that in addition
to 2T-physics there are other cherished symmetries in 1T field theory that
require the same structure, so this is not just an isolated weird field
theory. It was later noted that Weyl-symmetric supergravity\cite{IB-sugra}%
\cite{ferraraKallosh}\cite{BST}, as well as usual supergravity \cite{Weinberg}%
, also predict a similar sign-changing structure due to the Kaehler potential,
but this was swept under the rug in investigations of supergravity
\cite{Weinberg}. Thus, the possibility of a sign flip from gravity to
antigravity, that geodesically completes the spacetime, remained unknown until
the work in \cite{BC}-\cite{barsArayaBH}. The sign-changing feature of the
curvature term is an essential part of geodesic completeness in both spacetime
as well as in field space \cite{BC}-\cite{barsArayaBH}. For answers to
questions raised about unitarity or instability due to this sign flip see
\cite{barsJames}. In short, by now there remains no concerns about unitarity,
instability or the physical meaning of this setup, although more work is
welcome to better understand the interesting physics as indicated in
\cite{barsJames}.

A new feature introduced formally in the current paper (carried out casually
in \cite{BC}-\cite{barsArayaBH}) is how to take into account the smoothing
effects of a quantum theory of gravity as part of an effective field theory
for cosmology (EFTC). The EFTC would also be applicable to black holes, black
strings etc. \cite{barsArayaJames}. Although currently there is no universally
accepted theory of quantum gravity (QG), one of its universally expected
features is that gravitational singularities are softer or even possibly
non-existent in a successful QG. Assuming that this softer behavior is true in
principle, in attempts to capture general effects of QG in the form of an
EFTC, the effective theory would be physically wrong if the EFTC is too
singular. In an EFTC that is compatible with the smoother behavior of QG there
should be some restriction on which singular curvatures (or their powers) may
appear in the equations of motion when it is being applied close to
singularities\footnote{For example, string theory makes definite predictions
of higher curvature terms. Those are applicable only at low energies, and not
at all close to the singularities. At the Planck scale string theory provides
a totally different and non-singular description of the physics, but this is
not yet well understood. In any case, the high curvature terms are absent near
gravitational singularities.}. One reasonable way to insure this, is the
following proposal which is based on some past success: namely, define the
EFTC to be given by Eq.(\ref{action}) that includes the $R\left(  g\right)  $
term, with the additional condition of not admitting any other higher
curvature terms \textit{in the effective action} when it is being applied near
singularities. This restriction may seem ad-hoc, but the fact that, in
practice, it produces just the desired smoother mathematical properties of a
workable model including singularities may be taken as its temporary
justification. Namely, this EFTC turns out to be sufficiently well-behaved
mathematically, as well as being geodesically complete, despite the presence
of curvature singularities in the form of $R\left(  g\right)  $ and $R_{\mu
\nu}\left(  g\right)  $ that do appear in its equations of motion. Higher
non-trivial curvatures and/or their powers exists in the relevant manifolds
but \textit{these terms do not appear in the action or equations of motion}
derived from the proposed EFTC. Thanks to the underlying Weyl symmetry that is
still present, and that can transform curvatures to less singular expressions
in various gauges, the singular terms turn out to be mathematically manageable
in solving equations, computing \textit{gauge invariant physical quantities},
and establishing geodesic completeness, as already demonstrated amply in
\cite{BC}-\cite{barsArayaBH}. More along these lines will become apparent in
the remainder of this paper.

\section{Geodesically complete mini-superspace \label{superspace}}

The Friedmann equation, as parametrized in the context of the $\Lambda$CDM
model \cite{cosmReview}-\cite{riess} provides an approximate phenomenological
parametrization of the evolution of the universe in terms of some constant
\textit{dimensionless} measured parameters $\Omega_{i}$
\begin{equation}
\frac{H^{2}\left(  x^{0}\right)  }{H_{0}^{2}}=\Omega_{\Lambda}+\frac
{\Omega_{K}}{a_{E}^{2}\left(  x^{0}\right)  }+\frac{\Omega_{m}}{a_{E}%
^{3}\left(  x^{0}\right)  }+\frac{\Omega_{r}}{a_{E}^{4}\left(  x^{0}\right)
}+\frac{\Omega_{\sigma}+\Omega_{\alpha}}{a_{E}^{6}\left(  x^{0}\right)
}+\cdots,\; \label{Omegas}%
\end{equation}
where $a_{E}\left(  x^{0}\right)  $ is the scale factor in the
\textit{Einstein frame}, $H\left(  x^{0}\right)  $ is the Hubble parameter,
$H_{0}$ is the Hubble constant, the $\left[  \Omega_{\Lambda},\Omega
_{K},\Omega_{m},\Omega_{r},\Omega_{\sigma},\Omega_{\alpha}\right]  $ are
associated to the energy densities per unit volume respectively for [dark
energy, curvature, massive matter (dark and baryonic), radiation (massless
relativistic matter), scalar field, anisotropy]. According to data,
$\Omega_{\Lambda}=0.692\pm0.012$, $\Omega_{m}=0.308\pm0.012$ show that dark
energy and dark matter dominate the energy balance today (i.e. when
$a_{E}\left(  x_{today}^{0}\right)  =1$). Radiation is small such that
$\Omega_{m}+\Omega_{r}\simeq0.31,$ $\Omega_{K}=\left(  1-\sum_{i\neq K}%
\Omega_{i}\right)  =0.0002\pm\ 0.0026$ is computed from all the other
$\Omega_{i},$ finally $\left(  \Omega_{\sigma},\Omega_{\alpha}\right)  $ are
no greater than the error bars set on the other parameters.

As the universe expands $a_{E}\left(  x^{0}\right)  \rightarrow\infty,$
$\Omega_{\Lambda}$ will dominate the future accelerated expansion of the
universe. On the other hand, in the early universe, as $a_{E}\left(
x^{0}\right)  \rightarrow0,$ no matter how small the parameters $\left(
\Omega_{\sigma},\Omega_{\alpha}\right)  $ may be, the dominant term is
$\left(  \Omega_{\sigma}+\Omega_{\alpha}\right)  a_{E}^{-6}\left(
x^{0}\right)  $, and next are the terms in (\ref{Omegas}) in reverse order,
with $\Omega_{\Lambda}$ the least influential. Hence a scalar field and
anisotropy combined denominate the degrees of freedom that govern the
evolution of the universe close to cosmological singularities (big bang, big
crunch), and these cannot be neglected in any approach that attempts to
understand the very beginning. The typical cosmologist gives up at that point,
however in this paper I will show that persisting in this study leads to
unique initial conditions.

I emphasize that Eq.(\ref{Omegas}) \textit{ for }$a_{E}$\textit{ is in the
Einstein frame}. The spacetime in this frame is geodesically incomplete but,
via local scale (Weyl) invariance, it can be extended to the complete
spacetime shown in Fig.1, where the era after the big bang described by
Eq.(\ref{Omegas}), occupies only the patch labelled as the future quadrant II,
as explained below.
\begin{center}
\includegraphics[
height=1.7149in,
width=1.7149in
]%
{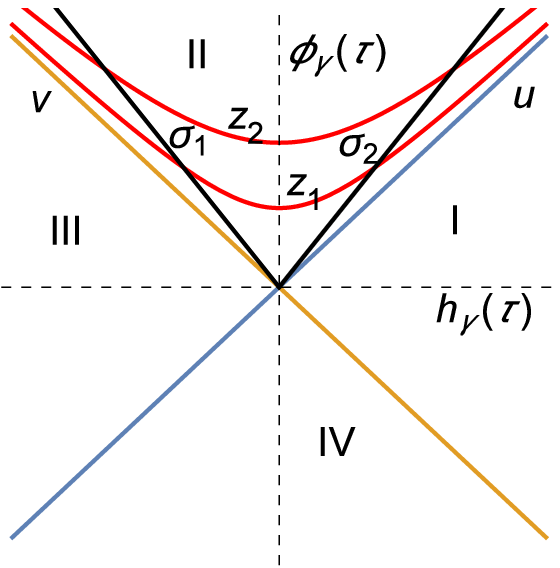}%
\\
Fig,(1){\protect\scriptsize  - Minkowski }$\left(  \phi_{\gamma},h_{\gamma
}\right)  ~${\protect\scriptsize  versus Rindler }$\left(  z,\sigma\right)
~${\protect\scriptsize  coordinates. In region II, the parabolas are at fixed
values of }$z${\protect\scriptsize , ~}$0<z_{1}<z_{2}<\infty,~$%
{\protect\scriptsize  and the rays are at fixed values of }$\sigma
${\protect\scriptsize , }$-\infty<\sigma_{1}<\sigma_{2}<\infty.~$%
{\protect\scriptsize  Similarly in regions I-IV.}%
\end{center}

The EFTC model in (\ref{action}) describes the evolution of the universe
consistently with (\ref{Omegas}), but in more detail, in terms of the
so-called \textquotedblleft mini-superspace\textquotedblright\ degrees of
freedom. These consist of the scale factor $a\left(  \tau\right)  ,$
anisotropy degrees of freedom $\alpha_{1,2}\left(  \tau\right)  $ in the
metric below (Bianchi I or VIII or IX), and the Higgs field $h\left(
\tau\right)  $ in the unitary gauge $H=\left(  0,h/\sqrt{2}\right)  ,$%
\begin{equation}%
\begin{array}
[c]{c}%
ds^{2}=a^{2}\left(  \tau\right)  \left(  -\left(  d\tau\right)  ^{2}%
e^{2}\left(  \tau\right)  +ds_{3}^{2}\right) \\
ds_{3}^{2}=e^{2\alpha_{1}\left(  \tau\right)  }\left(  e^{2\sqrt{3}\alpha
_{2}\left(  \tau\right)  }d\sigma_{x}^{2}+e^{-2\sqrt{3}\alpha_{2}\left(
\tau\right)  }d\sigma_{y}^{2}\right)  +e^{-4\alpha_{1}\left(  \tau\right)
}d\sigma_{z}^{2}.
\end{array}
\label{metric}%
\end{equation}
Then the mini-superspace action, $S_{\text{mini}}=\int d\tau\mathcal{L}%
_{\text{mini}},$ follows directly \cite{BCTsolutions} from the EFTC in
(\ref{action}) by dimensional reduction, keeping only the $\tau$-dependence of
fields%
\begin{equation}%
\begin{array}
[c]{c}%
\mathcal{L}_{\text{mini}}=\frac{1}{2e}\left[  -\left(  \partial_{\tau}\left(
\phi a\right)  \right)  ^{2}+\left(  \partial_{\tau}\left(  ha\right)
\right)  ^{2}+a^{2}\left(  \phi^{2}-h^{2}\right)  \left(  \left(
\partial_{\tau}\alpha_{1}\right)  ^{2}+\left(  \partial_{\tau}\alpha
_{2}\right)  ^{2}\right)  \right]  -\frac{e}{2}V\\
V=a^{4}\left(  \phi^{2}-h^{2}\right)  ^{2}f\left(  \frac{h}{\phi}\right)
+a^{2}\left(  \phi^{2}-h^{2}\right)  V_{K}\left(  \alpha_{1},\alpha
_{2}\right)  +\Omega_{c}%
\end{array}
\label{smini-general}%
\end{equation}

The different parts of the potential energy, $\Omega_{c},f\left(
h/\phi\right)  ,V_{K}\left(  \alpha_{1},\alpha_{2}\right)  $ come from the
following sources. The parameter $\Omega_{c}$ is related to the energy density
of all the conformally invariant matter described by the standard model term
$L_{SM}$ in (\ref{action}), when this matter is approximated by a
\textquotedblleft conformal dust\textquotedblright\ energy momentum tensor.
The $T_{00}$ component for $L_{SM}$ then has the form $\Omega_{c}/a^{4},$ just
like conformally invariant radiation appears in the Freedman equation
(\ref{Omegas}). So, the coefficient $\Omega_{c}=\Omega_{m}+\Omega_{r}%
\simeq0.31,$ includes dark matter, baryonic matter, as well as radiation. The
Higgs potential that appears in (\ref{action}) is written as, $V\left(
\phi,h\right)  =\left(  \phi^{2}-h^{2}\right)  ^{2}f\left(  \frac{h}{\phi
}\right)  ,$ and the anisotropy potential that arises from the metric
(\ref{metric}) is written as $\left(  \phi^{2}-h^{2}\right)  V_{K}\left(
\alpha_{1},\alpha_{2}\right)  ,$ where $V_{K}\left(  \alpha_{1},\alpha
_{2}\right)  $ was computed by Misner \cite{misner}. These are given
by\footnote{In (\ref{smini-general}) I choose units such that, the time
parameter $\tau$ is the conformal time $x^{0}$ in (\ref{Omegas}) rescaled by
the Hubble time, $\tau\equiv H_{0}x^{0};$ the dimensionful scalar degrees of
freedom $\phi,h$ in (\ref{action}) are rescaled by a factor of $\phi_{0}%
=\sqrt{12/16\pi G}$ defined in (\ref{universal}), $\left(  \phi,h\right)
=\phi_{0}\left(  \bar{\phi},\bar{h}\right)  $, so that the corresponding
symbols appearing in the cosmological analysis below are the dimensionless
$\left(  \bar{\phi},\bar{h}\right)  $. However, to avoid a proliferation of
symbols, instead of $\left(  \bar{\phi},\bar{h}\right)  $ the same symbols
$\left(  \phi,h\right)  $ will be understood to mean $\left(  \bar{\phi}%
,\bar{h}\right)  $ when there is no confusion. Similarly, the dimensionless
anisotropy degrees of freedom $\alpha_{1,2}$ in (\ref{metric}) are rescaled by
$\phi_{0}$ as compared to previous publications \cite{BC}-\cite{barsArayaBH}.
With this choice of units the mini-superspace action below contains the same
dimensionless parameters $\Omega_{i}$ that appear in the phenomenological
parametrization (\ref{Omegas}) of the Friedmann equation. \label{units}}
\begin{equation}%
\begin{array}
[c]{l}%
f\left(  h/\phi\right)  \equiv2\frac{\Omega_{\lambda}\left(  \left(
h/\phi\right)  ^{2}-w^{2}\right)  ^{2}+\Omega_{\Lambda}}{\left(  1-\left(
h/\phi\right)  ^{2}\right)  ^{2}},\left\{
\begin{array}
[c]{l}%
\Omega_{\Lambda}=\frac{\lambda^{\prime}}{4}\frac{3}{4\pi}\left(  \frac
{m_{P}t_{P}}{H_{0}t_{P}}\right)  ^{2}\simeq0.692\\
\Omega_{\lambda}=\frac{\lambda}{4}\frac{3}{4\pi}\left(  \frac{m_{P}t_{P}%
}{H_{0}t_{P}}\right)  ^{2}\simeq10^{120}%
\end{array}
\right.  ,\\
V_{K}\left(  \alpha_{1},\alpha_{2}\right)  \equiv\frac{k\left\vert \Omega
_{K}\right\vert }{4k-1}\left(
\begin{array}
[c]{c}%
e^{-8\alpha_{1}}+4e^{4\alpha_{1}}\sinh^{2}\left(  2\sqrt{3}\alpha_{2}\right)
\\
-4ke^{-2\alpha_{1}}\cosh\left(  2\sqrt{3}\alpha_{2}\right)
\end{array}
\right)  ,~\left\vert \Omega_{K}\right\vert \simeq0.0002,
\end{array}
\label{fAndVK}%
\end{equation}
where $m_{P},t_{P}$ are the Planck mass and time. In $V_{K}\left(  \alpha
_{1},\alpha_{2}\right)  $ the parameter $k=\left(  0,-1,+1\right)  $ is used
to distinguish the 3-dimensional anisotropic (flat, open, closed)-metrics,
Bianchi I,VIII, IX respectively. Note that, in the chosen units explained in
footnote (\ref{units}), $\Omega_{\lambda}\sim10^{120}$ is huge. However, this
term in the Higgs potential is suppressed because, just after the electroweak
phase transition (EW), the Higgs sits at the minimum of its potential,
$\left\vert h\left(  \tau\right)  /\phi\left(  \tau\right)  \right\vert
\rightarrow$ $\left\vert h_{0}/\phi_{0}\right\vert =w\sim10^{-17},$ during
most of the cosmological evolution of the universe.

\subsection{Mini Weyl symmetry, gauges and transformations among them}

$S_{\text{mini}}$ is invariant under local rescaling (Weyl) transformations
using the arbitrary time dependent gauge parameter $\lambda\left(
\tau\right)  ,$ namely $a\rightarrow\lambda^{-1}a,\;\phi\rightarrow\lambda
\phi,\;h\rightarrow\lambda h,\;\alpha_{1,2}\rightarrow\alpha_{1,2}.$ There are
three gauge dependent mini-superspace degrees of freedom $\left(
a,\phi,h\right)  $ while $\alpha_{1,2}$ are scale invariant. Other scale
invariants include $\left(  a\phi,ah,h/\phi\right)  .$ One may choose a Weyl
gauge in which some combination of $\left(  a,\phi,h\right)  $ is gauge fixed
for all $\tau$.

The \textquotedblleft$\gamma$-gauge\textquotedblright\ is defined by setting
the scale factor to $1$ for all $\tau$, $a_{\gamma}\left(  \tau\right)  =1,$
while $\phi_{\gamma}\left(  \tau\right)  ,h_{\gamma}\left(  \tau\right)  $
along with $\alpha_{1,2}\left(  \tau\right)  $ are the remaining dynamical
degrees of freedom. The label $\gamma$ emphasizes that the degrees of freedom
are defined in this gauge, however $\left(  \phi_{\gamma},h_{\gamma}\right)  $
are actually gauge invariants since $\left(  a\phi,ah\right)  =\left(
1\phi_{\gamma},1h_{\gamma}\right)  $\footnote{The $\gamma$-gauge is also
available in the full spacetime $x^{\mu}.$ It amounnts to fixing the
determinant of the metric $g_{\mu\nu}\left(  x^{\mu}\right)  $ to one for all
$x^{\mu}$, i,e, $\left(  -g\left(  x^{\mu}\right)  \right)  =1.$ So the
$\gamma$-gauge may also be called the uni-modular gauge for gravity.}$.$ As
will be clarified below, $\left(  \phi_{\gamma},h_{\gamma}\right)  $ turn out
to be global degrees of freedom that cover all the patches of the geodesically
complete mini-superspace partly shown in Fig.1. It is useful to define
$z\left(  \tau\right)  \equiv\left(  \phi_{\gamma}^{2}-h_{\gamma}^{2}\right)
$ and the sign$\left(  z\left(  \tau\right)  \right)  \equiv\varepsilon
_{z}\left(  \tau\right)  .$ The sign of $\left(  \phi^{2}-h^{2}\right)  $
cannot be changed under local Weyl rescalings, therefore $\varepsilon_{z}%
=$sign$\left(  \phi_{\gamma}^{2}-h_{\gamma}^{2}\right)  =$sign$\left(
\phi^{2}-h^{2}\right)  $ is gauge invariant, and $\varepsilon_{z}\left(
\tau\right)  =\pm1$ distinguishes between gravity/antigravity sectors at any
given $\tau$ as seen from Eq.(\ref{action}).

By contrast, the Einstein frame with its own $a_{E}\left(  \tau\right)  $ that
appears in phenomenological equations such as (\ref{Omegas}), emerges in the
\textquotedblleft E-gauge\textquotedblright\ which is defined by, $\left(
\phi_{E}^{2}-h_{E}^{2}\right)  =\left(  \phi_{E}-h_{E}\right)  \left(
\phi_{E}+h_{E}\right)  =\varepsilon_{z}\left(  \tau\right)  ,$ for all $\tau,$
and parametrized by $\left(  \phi_{E}+h_{E}\right)  =\pm^{\prime}%
e^{\sigma\left(  \tau\right)  },$ and $\left(  \phi_{E}-h_{E}\right)
=\pm^{\prime}\varepsilon_{z}e^{-\sigma\left(  \tau\right)  },$ where
$\pm^{\prime}$ is an additional set of signs that distinguish various regions
in Fig.1. The traditional Einstein-Hilbert theory corresponds to taking only
the patch $\left(  \phi_{E}+h_{E}\right)  >0$ and $\left(  \phi_{E}%
-h_{E}\right)  >0,$ which corresponds to $\pm^{\prime}\rightarrow+$ and also
$\varepsilon_{z}\left(  \tau\right)  \rightarrow+1,$ so that the conventional
theory is defined in the geodesically incomplete future quadrant shown in Fig.1.

By comparing gauge invariants in these two gauges, such as $a^{2}\left(
\phi^{2}-h^{2}\right)  =a_{E}^{2}\varepsilon_{z}=1\left(  \phi_{\gamma}%
^{2}-h_{\gamma}^{2}\right)  =z$, one learns
\begin{equation}
a_{E}^{2}=\left\vert z\right\vert =\left\vert \phi_{\gamma}^{2}-h_{\gamma}%
^{2}\right\vert . \label{z}%
\end{equation}
So the $a_{E}$ in the Friedmann equation is $a_{E}\left(  \tau\right)
=+\left\vert z\left(  \tau\right)  \right\vert ^{1/2}=+\left\vert \phi
_{\gamma}^{2}-h_{\gamma}^{2}\right\vert ^{1/2},$ noting that $z$ can be
positive (gravity sectors II ad IV in Fig.1) or negative (antigravity sectors
I\&III in Fig.1). From (\ref{z}) it is clear that the singularity in the
Einstein frame, $a_{E}^{2}=0,$ occurs only when $\left(  \phi_{\gamma}%
^{2}-h_{\gamma}^{2}\right)  $ vanishes, but when this vanishes in the $\gamma
$-gauge, $\left(  \phi^{2}-h^{2}\right)  $ in any gauge must also vanish since
the sign of this quantity is Weyl gauge invariant. The same argument holds for
all gravitational singularities (including black holes) in the Einstein frame,
hence these occur precisely when the coefficient of $R$ in the original Weyl
invariant action (\ref{action}) changes sign.

Similarly, by considering another set of gauge invariants, $\left(
a\phi,ah\right)  =\left(  a_{E}\phi_{E},a_{E}h_{E}\right)  =\left(
\phi_{\gamma},h_{\gamma}\right)  ,$ one finds the following transformation
between the global coordinates $\left(  \phi_{\gamma},h_{\gamma}\right)  $ and
the patchy E-frame coordinates $\left(  z,\sigma\right)  $, both sets being
Weyl invariants,
\begin{equation}%
\begin{array}
[c]{l}%
u=\phi_{\gamma}+h_{\gamma}=\pm^{\prime}\sqrt{\left\vert z\right\vert
}e^{\sigma},\;v=\phi_{\gamma}-h_{\gamma}=\pm^{\prime}\sqrt{\left\vert
z\right\vert }e^{-\sigma}\varepsilon_{z},\;-\infty<\phi_{\gamma},h_{\gamma
}<\infty,\\
z=\phi_{\gamma}^{2}-h_{\gamma}^{2}=uv,\;\sigma=\frac{1}{2}\ln\left\vert
\frac{\phi_{\gamma}+h_{\gamma}}{\phi_{\gamma}-h_{\gamma}}\right\vert =\frac
{1}{2}\ln\left\vert \frac{u}{v}\right\vert ;\;\;-\infty<z,\sigma<\infty.
\end{array}
\label{coordTransf}%
\end{equation}

For low energy physics, one should also keep track of the c-gauge, $\phi
_{c}\left(  x^{\mu}\right)  =\phi_{0}=1$ (in the units of footnote
\ref{units}), that was used to identify the universal constants
(\ref{universal}) and all low energy physics degrees of freedom$.$ Using the
Weyl gauge invariants $h/\phi$ and $a\phi$ one obtains $h/\phi=h_{E}/\phi
_{E}=h_{\gamma}/\phi_{\gamma}=h_{c}/1$ and $a\phi=a_{E}\phi_{E}=1\phi_{\gamma
}=a_{c}1.$ Hence the Weyl invariant low energy Higgs field $h_{c}$ and scale
factor $a_{c}$ are written in terms of the Weyl invariant cosmologically
global fields $\left(  \phi_{\gamma},h_{\gamma}\right)  ,$ and the Weyl
invariant patchy fields $\left(  z,\sigma\right)  $ of the E-gauge (related to
$a_{E}$ used in cosmological phenomenology as in (\ref{Omegas})), as follows
\begin{equation}
h_{c}=\frac{h_{\gamma}}{\phi_{\gamma}}=\frac{\varepsilon_{z}e^{2\sigma}%
-1}{\varepsilon_{z}e^{2\sigma}+1},\;a_{c}=\phi_{\gamma}=\pm^{\prime}\frac
{1}{2}\sqrt{\left\vert z\right\vert }\left(  e^{\sigma}+e^{-\sigma}%
\varepsilon_{z}\right)  . \label{Higgs}%
\end{equation}
Note that $\left(  h_{c},a_{c}\right)  $ are also global variables (i.e. not
patchy). At the observed low energies in today's era, $\varepsilon_{z}\left(
\tau\right)  =+1,$ in the future patch $\pm^{\prime}\rightarrow+,$ in Fig.1,
we have $\sigma\simeq h_{c}\simeq\frac{240\text{ GeV}}{10^{19}\text{ GeV}}%
\lll1$ and $a_{E}\left(  \tau\right)  \simeq a_{c}\left(  \tau\right)
=\phi_{\gamma}\left(  \tau\right)  .$ However, cosmologically none of these
quantities are small or close to each other numerically, so their distinct
meanings as given in (\ref{coordTransf},\ref{Higgs}) should be kept in mind
when discussing physics at various energy regimes and various cosmological eras.

The transformation of coordinates displayed in (\ref{coordTransf}) is
precisely the same as the transformation between 2-dimensional flat Minkowski
coordinates $\left(  \phi_{\gamma},h_{\gamma}\right)  $ and \textit{extended}
Rindler coordinates $\left(  z,\sigma\right)  $ as used recently in
\cite{barsAraya-rindler}, but now understood as part of the degrees of freedom
in mini-superspace
\begin{equation}
ds_{\text{mini}}^{2}=-dudv=-d\phi_{\gamma}^{2}+dh_{\gamma}^{2}=-\left(
4z\right)  ^{-1}dz^{2}+\left(  z\right)  d\sigma^{2}. \label{rind}%
\end{equation}
As shown in Fig.1, the $\gamma$-frame $\left(  \phi_{\gamma},h_{\gamma
}\right)  $ or $\left(  u,v\right)  $ cover globally all four quadrants of
extended Rindler space (see \cite{barsAraya-rindler} for more detail) with an
unambiguous identification of time-like $\left(  \phi_{\gamma}\right)  $ and
space-like $\left(  h_{\gamma}\right)  $ coordinates is a geodesically
complete mini-superspace. The curvature singularity that occurs in the
E-frame, when $a_{E}^{2}=0,$ corresponds to $z=0$ which translates to either
$u=0$ or $v=0$ in the flat global space of Eq.(\ref{rind}). So the
cosmological bang or crunch singularities of the E-frame can occur only at the
horizons of the $\gamma$-frame that form the boundaries of the four Rindler
quadrants in Fig.1.

This globally flat 2D-Minkowski geometry is the intrinsic geometrical property
of the scale invariant mini-superspace in any frame, including the
geodesically completed E-frame written in terms of $z$ as in (\ref{rind}).
This is the underlying reason for how it is possible to go through
cosmological singularities - that amount to horizons in global coordinates -
to complete geodesics in complete field space in mini-superspace, as well as
space-time $x^{\mu},$ as explored extensively in \cite{BC}-\cite{barsArayaBH}.

\section{Quantum wavefunction - 1 \label{QM1}}

The mini-superspace action (\ref{smini-general}) can now be expressed in the
$\gamma$-gauge in terms of the Weyl invariant $\left(  \phi_{\gamma}%
,h_{\gamma}\right)  $ degrees of freedom by setting $a_{\gamma}=1.$ From this
point on, the $\gamma$ label will be suppressed for simplicity and $\left(
\phi,h\right)  $ will be understood to mean $\left(  \phi_{\gamma},h_{\gamma
}\right)  $ when there is no confusion.
\begin{equation}%
\begin{array}
[c]{l}%
\mathcal{L}_{\text{mini}}^{\gamma}=\frac{1}{2e}\left[  -\dot{\phi}^{2}+\dot
{h}^{2}+\left(  \phi^{2}-h^{2}\right)  \left(  \dot{\alpha}_{1}^{2}%
+\dot{\alpha}_{2}^{2}\right)  \right]  -\frac{e}{2}V,\\
V=\left(  \phi^{2}-h^{2}\right)  ^{2}f\left(  \frac{h}{\phi}\right)  +\left(
\phi^{2}-h^{2}\right)  V_{K}\left(  \alpha_{1},\alpha_{2}\right)  +\Omega
_{c},\\
\mathcal{H}=\left[  -\pi_{\phi}^{2}+\pi_{h}^{2}+\frac{1}{\phi^{2}-h^{2}%
}\left(  \pi_{1}^{2}+\pi_{2}^{2}\right)  +V\right]  =0.
\end{array}
\label{Sminifh}%
\end{equation}
The last line is the constraint that follows from the $e$ equation of motion,
$\mathcal{H}=\partial S_{\text{mini}}/\partial e\left(  \tau\right)  =0.$ This
is the vanishing Hamiltonian $\mathcal{H}$ expressed in terms of the canonical
momenta $\left(  \pi_{\phi}=-\dot{\phi}/e,\cdots,\pi_{2}=\dot{\alpha}_{2}%
^{2}/e\right)  $ for any $e\left(  \tau\right)  .$ Note that there is no need
to gauge fix the lapse function $e\left(  \tau\right)  $ due to $\tau$
reparametrization symmetry since the properties of the canonical phase space
in $\mathcal{H}$ is insensitive to a gauge choice for $e\left(  \tau\right)
.$ Straightforward quantization rules applied to this system, and applying the
constraint on physical states, $\mathcal{H}\Psi=0,$ produces the Wheeler
deWitt equation (WdWe) that follows from $\mathcal{L}_{\text{mini}}^{\gamma}%
$,
\begin{equation}
\left[  \partial_{\phi}^{2}-\partial_{h}^{2}-\frac{1}{\phi^{2}-h^{2}}\left(
\partial_{1}^{2}+\partial_{2}^{2}\right)  +V\right]  \Psi=0. \label{WdWfh}%
\end{equation}
There is no ambiguity of quantum ordering problems in the quantum phase space
as it appears in $\mathcal{H}$ above in contrast to other choices of
mini-superspace parametrizations such as $\left(  z,\sigma,\alpha_{1}%
,\alpha_{2}\right)  $. Choosing the $\left(  \phi,h\right)  $ global
coordinates (which are the ones naturally appearing in the full action
(\ref{action})), as the preferred degrees of freedom in the definition of the
quantum theory, resolves once and for all this long-standing annoying quantum
ambiguity \cite{hawkingHartle}\cite{halliwell-q}\footnote{The ambiguity in the
ordering prescription proposed in \cite{halliwell-q} is to write the kinetic
terms in (\ref{WdWfh}) in the form of the Klein-Gordon operator with an added
curvature term with an unknown $\xi$ coefficient, $\left(  \nabla^{2}+\xi
R\left(  g\right)  +V\right)  \Phi=0,$ where $\nabla^{2}\Phi=\left(
-g\right)  ^{-1/2}\partial_{\mu}\left(  \left(  -g\right)  ^{1/2}g^{\mu\nu
}\partial_{\nu}\Phi\right)  .$ In the current case the metric is conformally
flat, $ds^{2}=-d\phi^{2}+dh^{2}+\left(  \phi^{2}-h^{2}\right)  \left(
d\alpha_{1}^{2}+d\alpha_{2}^{2}\right)  =-dudv+uv\left(  d\alpha_{1}%
^{2}+d\alpha_{2}^{2}\right)  ,$ and its curvature is $R\left(  g\right)
=6\left(  \phi^{2}-h^{2}\right)  ^{-1}.$ In this expression replacing $\Phi$
by $\Phi=$ $\left(  \phi^{2}-h^{2}\right)  ^{-1/2}\Psi$ and also fixing
$\xi=-1/6$, reproduces precisely Eq.(\ref{WdWfh}) for $\Psi.$ This shows that
the straightforward no-need-to-order prescription applied to obtain
(\ref{WdWfh}) is in agreement with \cite{halliwell-q} but only when
$\xi=-1/6,$ indicating that the ambiguity in \cite{halliwell-q} is fully
resolved by the preferred quantum \textit{global coordinates}.}.

Having resolved the quantum ordering, the WdWe can now be rewritten in the
$\left(  z,\sigma\right)  $ basis in the Einstein frame (i.e. in terms of
$a_{E}$ used by phenomenologists) by using the coordinate transformation
(\ref{coordTransf}) and noting the non-trivial ordering that is uniquely
predicted in the $z$ variable, $\partial_{\phi}^{2}-\partial_{h}^{2}%
=4\partial_{u}\partial_{v}=4z\partial_{z}^{2}+4\partial_{z}-z^{-1}%
\partial_{\sigma}^{2},$ while the rest is straightforward. The WdWe in the
$\left(  z,\sigma\right)  $ basis is then manipulated to the following
non-relativistic Schr\"{o}dinger-type equation form
\begin{equation}
\left[  -\partial_{z}^{2}-\frac{1}{4z^{2}}\left(  1-\partial_{1}^{2}%
-\partial_{2}^{2}-\partial_{\sigma}^{2}\right)  -\frac{\Omega_{c}}{4z}%
-\frac{1}{4}V_{K}\left(  \alpha_{1},\alpha_{2}\right)  -\frac{z}{4}V\left(
\sigma,\varepsilon_{z}\right)  \right]  \left(  \sqrt{z}\Psi\right)  =0
\label{WdWezs}%
\end{equation}
where $V\left(  \sigma,\varepsilon_{z}\right)  =f\left(  h/\phi\right)  $
after using \ref{coordTransf}. The term $1/4z^{2}$ arises from rewriting
$\left(  4z\partial_{z}^{2}+4\partial_{z}\right)  \Psi=\sqrt{z}\left(
4\partial_{z}^{2}+z^{-2}\right)  \left(  \sqrt{z}\Psi\right)  .$

Thinking of $z$ as a \textquotedblleft time\textquotedblright\ variable,
(\ref{WdWezs}) can be viewed as a time-dependent Hamiltonian problem in
Schr\"{o}dinger-equation-type quantum mechanics for which well known
time-dependent methods exist to make progress and interpret the physics.
Nevertheless, this is a difficult partial differential equation in the
presence of the Higgs and anisotropy potentials $V\left(  \sigma
,\varepsilon_{z}\right)  $, $V_{K}\left(  \alpha_{1},\alpha_{2}\right)  ,$ so
numerical methods will be needed to analyze it fully. However, there is no
substitute for analytic approximations that can guide such numerical efforts.
This provides an incentive to look for circumstances that make it possible to
find approximate analytic methods to solve (\ref{WdWezs}). I suggest the
following approach.

It was noted following (\ref{fAndVK}) that, the large term $\Omega_{\lambda
}\sim10^{120}$ in the Higgs potential is suppressed because, just after the
electroweak phase transition (EW), the Higgs sits at the minimum of its
potential, $\left\vert h\left(  \tau\right)  /\phi\left(  \tau\right)
\right\vert \rightarrow$ $\left\vert h_{0}/\phi_{0}\right\vert =w\sim
10^{-17},$ during most of the cosmological evolution of the universe.
Furthermore before EW and close to the singularity $z\simeq0$ in the very
early universe, the Higgs and anisotropy potential terms in (\ref{WdWezs}) are
subdominant due to the factors of a vanishing $z.$ Therefore, the terms
involving $V_{K}\left(  \alpha_{1},\alpha_{2}\right)  ,V\left(  \sigma
,\varepsilon_{z}\right)  $ in (\ref{WdWezs}) can be neglected in the
computation of the wavefunction near $z\simeq0$ before EW, as well as well as
after EW because the Higgs settles down to the bottom of the potential.

Based on the comments in the previous paragraph, I observe that, as the system
moves away from singularities, the degrees of freedom $\vec{s}=\left(
\alpha_{1},\alpha_{2},\sigma\right)  $ will quickly descend to the ground
state in their respective potential energies, and stay there during most of
the evolution of the universe. This observation is consistent with general
physical behavior of degrees of freedom subjected to attractive time dependent
potentials, which is the case in the current problem. It is also consistent
both with cosmological data as well as the behavior of classical solutions of
these degrees of freedom as studied in the past, both analytically
\cite{BCTsolutions}-\cite{BCSTcomplete} and numerically \cite{BST-Higgs}. I
use these facts to devise the following 2-step strategy to approximate the
effects of the potentials in cosmological calculations. Briefly,

\begin{enumerate}
\item The first step of this strategy is an approximation that replaces the
functions $V\left(  \sigma,\varepsilon_{z}\right)  ,V_{K}\left(  \alpha
_{1},\alpha_{2}\right)  $ by constant values at their lowest energy
configuration
\begin{equation}
V_{K}\left(  \alpha_{1},\alpha_{2}\right)  \rightarrow-k\left\vert \Omega
_{K}\right\vert ,\;V\left(  \sigma,\varepsilon_{z}\right)  \rightarrow
2\Omega_{\Lambda}. \label{Vapprox}%
\end{equation}
In this step, (\ref{WdWezs}) turns into the following much simpler second
order ordinary differential equation that has analytic solutions,
\begin{equation}
\left(  -\partial_{z}^{2}+V\left(  z\right)  \right)  \left(  \sqrt{z}%
\Psi\right)  =0,\;\;V\left(  z\right)  =-\left(  \frac{\vec{p}^{2}+1}{4z^{2}%
}+\frac{\Omega_{c}}{4z}-\frac{\Omega_{K}}{4}+\frac{\Omega_{\Lambda}}%
{2}z\right)  . \label{WdWSch}%
\end{equation}
In this form $\Psi$ is taken to momentum space thus diagonalizing the operator
$\left(  -\partial_{1}^{2}-\partial_{2}^{2}-\partial_{\sigma}^{2}\right)
\rightarrow\vec{p}^{2},$ where $\vec{p}=\left(  p_{1},p_{2},p_{3}\right)  $
are the canonical conjugates to $\vec{s}=\left(  \alpha_{1},\alpha_{2}%
,\sigma\right)  .$ Note also in (\ref{WdWSch}) there is an accidental
SO$\left(  3\right)  $ symmetry that rotates the vectors $\left(  \vec{p}%
,\vec{s}\right)  .$ Then $\Psi_{\pm\left\vert \vec{p}\right\vert }\left(
z\right)  $ are the two linearly independent solutions of (\ref{WdWSch}). The
general solution is the superposition of the complete set of states in
momentum space, namely
\begin{equation}
\Psi\left(  z,\vec{s}\right)  =\int d^{3}pe^{-i\vec{p}\cdot\vec{s}}\left(
A_{+}\left(  \vec{p}\right)  \Psi_{+\left\vert \vec{p}\right\vert }\left(
z\right)  +A_{-}\left(  \vec{p}\right)  \Psi_{-\left\vert \vec{p}\right\vert
}\left(  z\right)  \right)  , \label{packet}%
\end{equation}
and this needs to be continuous in the geodesically complete superspace in
Fig.1. The latter is not trivial as discussed in section \ref{QM2}.

\item The second step of the strategy is to insure that the momenta $\vec{p}$
are limited in magnitude because these degrees of freedom will be sitting in
their ground state, so their kinetic energy cannot exceed the total energy of
the respective ground states. The limit set on the size of $p_{3}^{2}$ has the
physical interpretation of the cosmological parameter $\Omega_{\sigma}$ that
measures the energy density of the scalar field in the Friedmann equation
\ref{Omegas}. Similarly, the limit set on $\left(  p_{1}^{2}+p_{2}^{2}\right)
$ has the interpretation of $\Omega_{\alpha}$ that measures the energy density
of anisotropy. This limitation will be taken into account by requiring the
wavepacket coefficients to behave like Gaussians (or something of that form)
controlled by the parameters $\left(  \Omega_{\sigma},\Omega_{\alpha}\right)
$%
\begin{equation}
A_{\pm}\left(  \vec{p}\right)  \sim e^{-\left(  p_{1}^{2}+p_{2}^{2}\right)
/2\Omega_{\alpha}}e^{-p_{3}^{2}/2\Omega_{\sigma}}. \label{gaussian}%
\end{equation}

\end{enumerate}

Before an analytic solution of (\ref{WdWSch}), it is valuable to understand
the qualitative physical behavior of the wavepacket $\left(  \sqrt{z}%
\Psi\left(  z,\vec{s}\right)  \right)  $ by examining the potential $V\left(
z\right)  $ in (\ref{WdWSch}) that is plotted in Figs.2. The values of the
parameters $\left(  \left\vert \vec{p}\right\vert ,\Omega_{c},\Omega
_{K},\Omega_{\Lambda}\right)  $ in Fig.2 are not the measured ones, but are
taken in a range that pictorially emphasize the essential physical features of
the potential $V\left(  z\right)  $. In addition, Fig.3 is included for the
closely related potential (the solid curve), $\tilde{V}\left(  z\right)
=-(\frac{\vec{p}^{2}+1}{4z^{2}}+\frac{\Omega_{\Lambda}}{2}z),$ where
$\Omega_{c},\Omega_{K}$ are set to $0$. The parameters $\left\vert \vec
{p}\right\vert ,\Omega_{\Lambda}$ in $\tilde{V}$ are those in $V$ that are the
most dominant at $z=0$ (i.e. $\vec{p}^{2}+1$) and the most dominant at large
$z$ (i.e. $\Omega_{\Lambda}$). The dashed curve in Fig.3 corresponds to
$V\left(  z\right)  $ including all the parameters$,$ so comparing the solid
and dashed curves in Fig.3 shows the qualitative effect of the additional
parameters $\left(  \Omega_{c},\Omega_{K}\right)  $. I remark that the leading
terms of the basic solutions $\Psi_{\pm\left\vert \vec{p}\right\vert }\left(
z\right)  ,$ as $z\rightarrow0^{\pm}$ and as $z\rightarrow\pm\infty,$ that
will be needed shortly, are the same for $V$ or $\tilde{V}$.

\begin{center}%
{\parbox[b]{2.6117in}{\begin{center}
\includegraphics[
height=1.6207in,
width=2.6117in
]%
{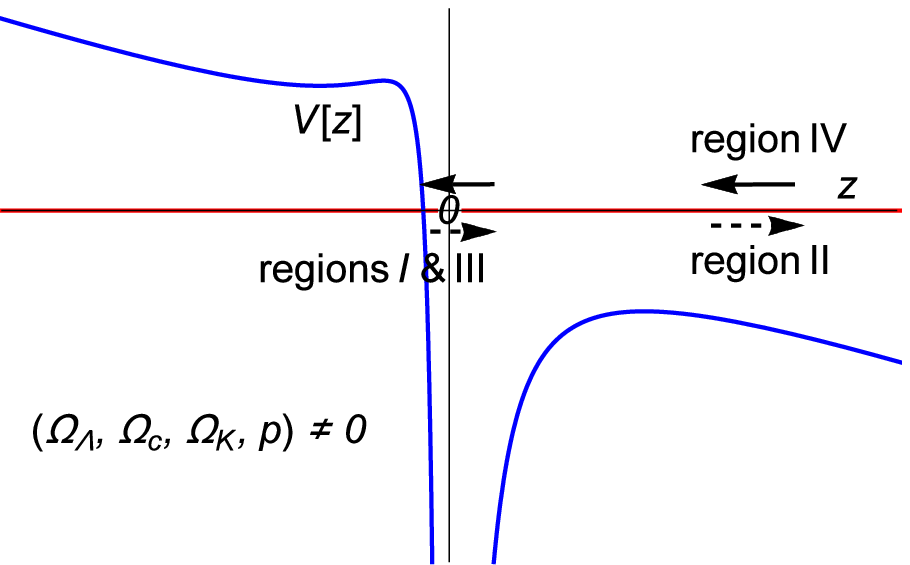}%
\\
{\protect\scriptsize Fig.2 - }$V\left(  z\right)  $ {\protect\scriptsize  with
all parameters }$\Omega_{i}\neq0.$%
\end{center}}}%
~~%
{\parbox[b]{2.6515in}{\begin{center}
\includegraphics[
height=1.6457in,
width=2.6515in
]%
{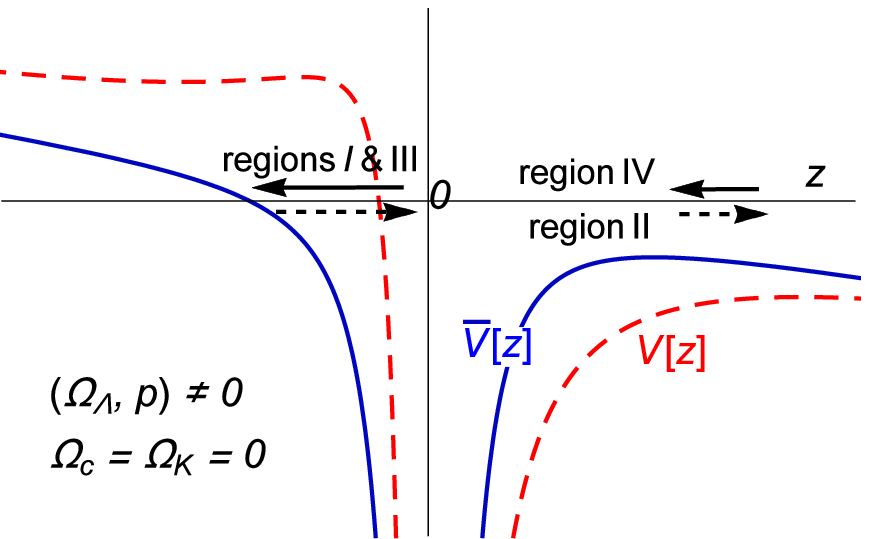}%
\\
{\protect\scriptsize Fig.3 - }$\tilde{V}\left(  z\right)  .$%
{\protect\scriptsize only dominant .}$\left\vert {\protect\scriptsize \vec{p}%
}\right\vert ,{\protect\scriptsize \Omega}_{{\protect\scriptsize \Lambda}}.$%
\end{center}}}%

\end{center}

The physical behavior of the wavepacket $\left(  \sqrt{z}\Psi\right)  $ can be
read off directly from Fig.2 or 3 as follows. The Schr\"{o}dinger energy
eigenvalue in (\ref{WdWSch}) is zero, so the energy level of the corresponding
eigenstate coincides with the horizontal axis in Figs.2,3. The wavepacket
spreads from $z\sim+\infty$ as indicated by the left pointing arrow in
Figs.2,3. This corresponds to a contracting universe evolving from the
asymptotic past in quadrant IV in Fig.1. The wavepacket in this region must
then behave like an incoming scattering state oscillatory solution because the
energy level is higher than the potential. The wavepacket passes through the
singularity at $z=0$, meaning it propagates continuously through the past
horizons in Fig.1 where the universe experiences a big crunch, so it reaches
into the antigravity regions I\&III where $z<0$. The wavepacket cannot go deep
into antigravity in regions I\&III because of the potential barrier in
Figs.2,3, so it gets reflected toward $z=0,$ meaning it turns around within
regions I \& III and propagates toward to future horizons in Fig.1. The part
of the wavepacket that tunnels under the mountain in Figs.2,3 cannot be an
oscillatory solution and it must exponentially decay away as $z$ becomes more
negative - this is because the potential is higher than the total energy in
that region and the probability must vanish in the asymptotic parts in
antigravity regions I \& III. The reflected wavepacket is oscillatory, it
passes through the singularity at $z=0$ again to come back to $z>0$ as shown
by the right pointing arrows in Figs.2,3, meaning the universe propagates
through the future horizons with a big bang into region II in Fig.1. After
this, the wavepacket propagates to larger values of $z,$ meaning the universe
expands in the future region II in Fig.1.

The qualitative physical properties in this account will be encoded in the
analytic expressions for time dependent classical solutions for $\left(
z\left(  \tau\right)  ,\vec{s}\left(  \tau\right)  \right)  $ as well as in
analytic wavepackets $\Psi\left(  z,\vec{s}\right)  $ given in the following sections.

\section{Dynamical attractor and initial conditions \label{attractor}}

It is helpful to begin with the classical solution version of the physical
scenario at the end of last section. The classical action $S_{\text{mini}}$ is
then defined by inserting the approximation (\ref{Vapprox}) in (\ref{Sminifh}%
)
\begin{equation}
S_{\text{mini}}\simeq\int d\tau\left\{  \frac{1}{2e}\left[  -\frac{1}%
{4z}\left(  \partial_{\tau}z\right)  ^{2}+z\left(  \partial_{\tau}\vec
{s}\right)  ^{2}\right]  -\frac{e}{2}\left[  2\Omega_{\Lambda}z^{2}-\left\vert
\Omega_{K}\right\vert z+\Omega_{c}\right]  \right\}  . \label{Sclassical}%
\end{equation}
The equations of motion for $z\left(  \tau\right)  $ and $\vec{s}\left(
\tau\right)  $ that follow from this action are reduced to first order
differential equations
\begin{equation}
z\left(  \partial_{\tau}\vec{s}\right)  =\vec{p},\;\;\left(  \partial_{\tau
}z\right)  ^{2}=\left[  4\vec{p}^{2}+8\Omega_{\Lambda}z^{3}-4\left\vert
\Omega_{K}\right\vert ~z^{2}+4z\Omega_{c}\right]  . \label{eomsGen}%
\end{equation}
Here $\vec{p}$ is the canonical conjugate to $\vec{s}$ which is a constant
$\partial_{\tau}\vec{p}=0$ due to the Euler-Lagrange equations of motion that
follow from $S_{\text{mini}}.$ The first order differential equation of motion
for $z\left(  \tau\right)  $ is a rewriting of the constraint that follows
from $\partial S_{\text{mini}}/\partial e=0.$ This is a first integral of the
second order differential equation of motion for $z\left(  \tau\right)  $ that
can be derived from $S_{\text{mini}}$.

I have obtained the general solution of (\ref{eomsGen}) with arbitrary initial
conditions for $z\left(  \tau\right)  $ and $\vec{s}\left(  \tau\right)  .$
The reader can verify that $z\left(  \tau\right)  $ is given analytically in
terms of the doubly periodic JacobiCN$[u|m]$ elliptic function usually denoted
as cn$\left(  u|m\right)  $, as follows%
\begin{equation}%
\begin{array}
[c]{l}%
z\left(  \tau\right)  =-\left\vert z_{0}\right\vert +z_{2}\frac{1-\text{cn}%
\left(  \sqrt{8\Omega_{\Lambda}z_{2}}\tau\left\vert ~m\right.  \right)
}{1+\text{cn}\left(  \sqrt{8\Omega_{\Lambda}z_{2}}\tau\left\vert ~m\right.
\right)  },\;\;m\equiv\frac{1}{2}+\frac{\left\vert z_{0}\right\vert +z_{1}%
}{2z_{2}},\\
z_{1}\equiv\frac{1}{2}\left(  \left\vert z_{0}\right\vert +\frac{\Omega_{K}%
}{2\Omega_{\Lambda}}\right)  >0,\;z_{2}\equiv\sqrt{z_{0}^{2}+2\left\vert
z_{0}\right\vert z_{1}+\frac{\vec{p}^{2}}{2\Omega_{\Lambda}\left\vert
z_{0}\right\vert }}>\left(  z_{1}+\left\vert z_{0}\right\vert \right)  .\;
\end{array}
\label{zLambda}%
\end{equation}
Here $z_{0},z_{1},z_{2}$ are constants determined by $\left(  \vec{p}%
^{2},\Omega_{\Lambda},\Omega_{K},\Omega_{c}\right)  $ as given below. In
particular $z_{0}$ is the value of $z\left(  \tau\right)  $ where the bracket
in (\ref{eomsGen}) vanishes, $\left[  \cdots\right]  =0,$ which occurs at the
instant $\dot{z}\left(  \tau\right)  =0.$ Due to the $\tau$-translation
symmetry of this system, one may choose this instant to be $\tau=0.$ Note that
$z_{0}$ is negative since it corresponds to the location of the barrier in
Figs.2,3, so it is determined as the only real finite root of the cubic
equation $\left[  4\vec{p}^{2}+8\Omega_{\Lambda}z^{3}-4\left\vert \Omega
_{K}\right\vert ~z^{2}+4z\Omega_{c}\right]  =0.$ Thus, $z_{0}\equiv z\left(
\tau=0\right)  ,$ is $z\left(  \tau\right)  $ at the instant $\tau=0$ when it
reaches its most negative \textit{classical} (as opposed to quantum) value in
the antigravity regime. The explicit solution for the relevant root $z_{0}$ is
written as follows (the other two roots are complex because of the physical
values of the parameters $\left(  \vec{p}^{2},\Omega_{\Lambda},\Omega
_{K},\Omega_{c}\right)  $ given is section \ref{superspace}).%
\begin{equation}%
\begin{array}
[c]{l}%
z_{0}=-\frac{1}{6\Omega_{\Lambda}}\left(  \left(  R+R\cos\phi\right)
^{\frac{1}{3}}-\left(  R-R\cos\phi\right)  ^{\frac{1}{3}}-\Omega_{K}\right)
<0,\\
R\equiv\left[  \left(  \Omega_{K}^{3}-54\Omega_{\Lambda}^{2}\vec{p}%
^{2}-9\Omega_{\Lambda}\Omega_{c}\Omega_{K}\right)  ^{2}+\left(  6\Omega
_{\Lambda}\Omega_{c}-\Omega_{K}^{2}\right)  ^{3}\right]  ^{\frac{1}{2}},\\
\cos\phi\equiv\frac{1}{R}\left(  54\Omega_{\Lambda}^{2}\vec{p}^{2}%
+9\Omega_{\Lambda}\Omega_{c}\Omega_{K}-\Omega_{K}^{3}\right)  >0.
\end{array}
\label{a}%
\end{equation}
The reader can verify analytically that (\ref{zLambda}-\ref{a}) is the general
solution of (\ref{eomsGen}) by using properties of the Jacobi elliptic
functions, namely $\partial_{u}$cn$\left(  u|m\right)  =-$sn$\left(
u|m\right)  \times$dn$\left(  u|m\right)  ,\;$sn$^{2}\left(  u|m\right)
+$cn$^{2}\left(  u|m\right)  =1$ and dn$\left(  u|m\right)  =1-m$
sn$^{2}\left(  u|m\right)  .$

The exact solution for $\vec{s}\left(  \tau\right)  $ that follows from
$z\left(  \partial_{\tau}\vec{s}\right)  =\vec{p}$ is,
\begin{equation}%
\begin{array}
[c]{l}%
\vec{s}\left(  \tau\right)  =\vec{s}_{0}+\vec{p}\int_{0}^{\tau}\frac
{d\tau^{\prime}}{z\left(  \tau^{\prime}\right)  }=\vec{s}_{0}+\vec{p}%
\int_{-\left\vert z_{0}\right\vert }^{z\left(  \tau\right)  }\frac{dz^{\prime
}}{z^{\prime}\dot{z}^{\prime}}\\
\;\;\;\;=\vec{s}_{0}+\vec{p}\int_{-\left\vert z_{0}\right\vert }^{z\left(
\tau\right)  }\frac{dz^{\prime}}{z^{\prime}\sqrt{4\vec{p}^{2}+8\Omega
_{\Lambda}z^{\prime3}-4\left\vert \Omega_{K}\right\vert ~z^{\prime
2}+4z^{\prime}\Omega_{c}}}=\vec{s}_{0}+\vec{p}~f\left(  z\left(  \tau\right)
\right)  ,
\end{array}
\label{stau}%
\end{equation}
Here, $\vec{s}_{0}=\vec{s}\left(  \tau=0\right)  $ is an integration constant
chosen as the value of $\vec{s}\left(  \tau\right)  $ at the instant $\tau=0$
when $z\left(  \tau\right)  $ is most negative in the antigravity regime. The
positive square root is used in determining $\dot{z}$ from (\ref{eomsGen})
because of the choice of integration region $0<\tau$. The function $f\left(
z\left(  \tau\right)  \right)  $ that is common to all directions of $\vec{p}%
$, can be written explicitly in terms of the Jacobi$\Pi$ function and has the
same double periodicity properties as $z\left(  \tau\right)  .$ The result
$f\left(  z\left(  \tau\right)  \right)  $ is valid not only for $\tau>0$ but
also for $\tau<0$ because the solution for $z\left(  \tau\right)  $ is
symmetric under $\tau$-reversal, $z\left(  -\tau\right)  =z\left(
\tau\right)  $ as seen in the figures below.

These functions are plotted in figures 4,5,6 using non-realistic values of the
phenomenological parameters to emphasize the important physical features of
$\left(  z\left(  \tau\right)  ,\vec{s}\left(  \tau\right)  \right)  $. Fig.4
shows a universe $z\left(  \tau\right)  $ (recall $a_{E}^{2}\left(
\tau\right)  =\left\vert z\left(  \tau\right)  \right\vert $) that begins to
contract from infinite size $\left(  z\left(  -T_{+}/2\right)  \sim
\infty\right)  ,$ passes through zero size $\left(  z\left(  -T_{0}\right)
=0\right)  ,$ reaches a maximum size in antigravity $\left(  z\left(
0\right)  =-\left\vert z_{0}\right\vert \right)  ,$ contracts back to zero
size $\left(  z\left(  T_{0}\right)  =0\right)  $ and then expands up to
infinity $\left(  z\left(  T_{+}/2\right)  \sim\infty\right)  $. This
classical behavior of $z\left(  \tau\right)  $ is in line with the qualitative
description of a wavepacket given following Figs.2,3.

\begin{center}%
{\parbox[b]{2.0387in}{\begin{center}
\includegraphics[
height=1.2656in,
width=2.0387in
]%
{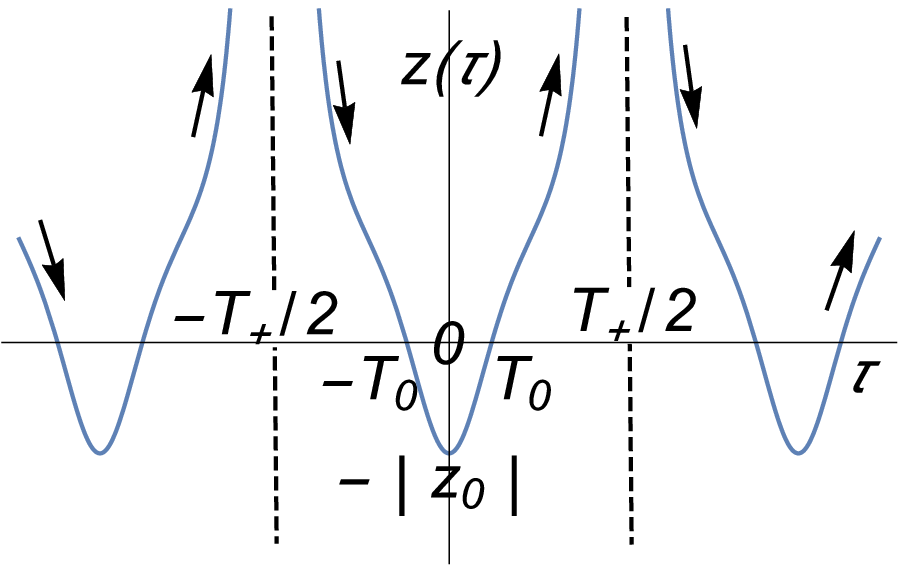}%
\\
{\protect\scriptsize Fig.4-}${\protect\scriptsize z(\tau)}$%
{\protect\scriptsize , crunch }${\protect\scriptsize \rightarrow}$
{\protect\scriptsize antigravity }${\protect\scriptsize \rightarrow}$
{\protect\scriptsize bang}%
\end{center}}}%
\ \
{\parbox[b]{1.9556in}{\begin{center}
\includegraphics[
height=1.2407in,
width=1.9556in
]%
{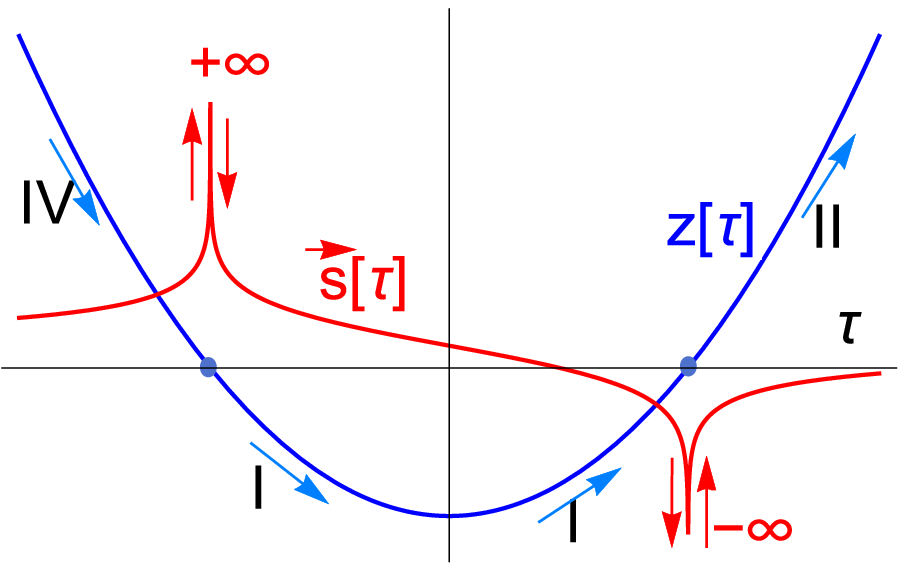}%
\\
{\protect\scriptsize Fig.5- }${\protect\scriptsize \vec{s}(\tau)}%
~${\protect\scriptsize  near }${\protect\scriptsize z(\tau)\simeq0}$%
\end{center}}}%
\ \ \
{\parbox[b]{1.7287in}{\begin{center}
\includegraphics[
height=1.3892in,
width=1.7287in
]%
{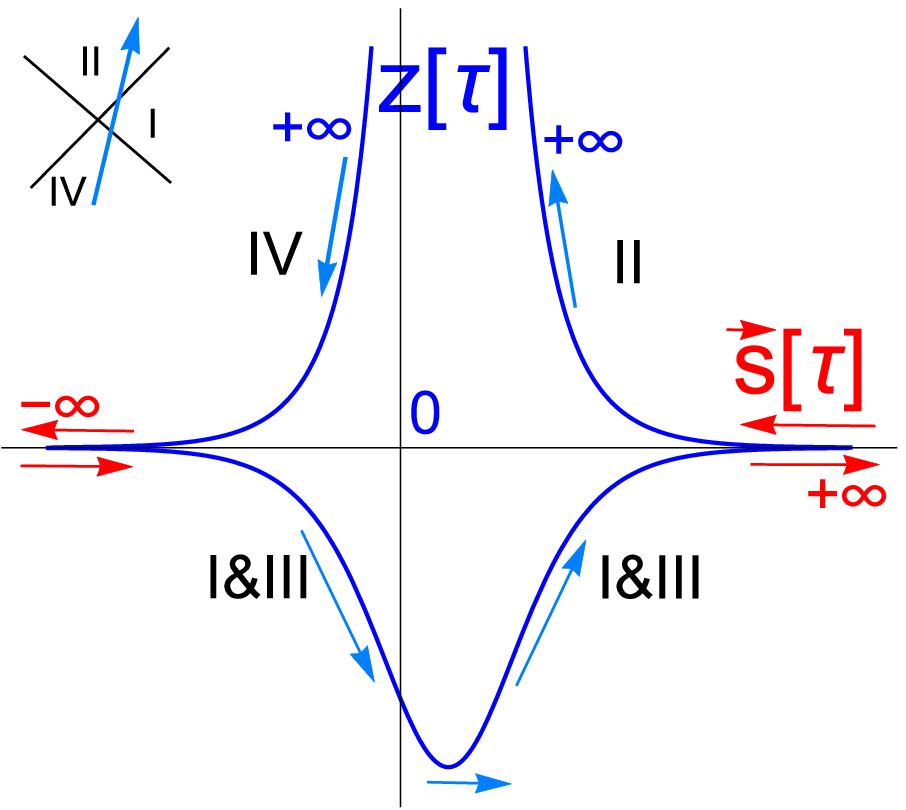}%
\\
{\protect\scriptsize Fig.6- Parametric (}${\protect\scriptsize \vec{s}(\tau
),}${\protect\scriptsize z(}${\protect\scriptsize \tau}$%
{\protect\scriptsize ))}%
\end{center}}}%

\end{center}

An interesting feature in Fig.4 is the periodicity seen in conformal time for
$z\left(  \tau+T_{+}\right)  =z\left(  \tau\right)  ;$ This is because of the
periodicity of the Jacobi elliptic function. Cosmic observers that use cosmic
$t$ as time (not conformal time $\tau$), can detect only one period of the
periodic Jacobi Elliptic functions. But it is intriguing that more generally
the plot in conformal time in Fig.4 shows a universe that gets renewed
periodically an infinite number of times, like a cyclic universe\footnote{This
is similar in spirit but different in detail than \cite{BST-Higgs} where
$\lambda^{\prime}$ was taken artificially negative to imitate tunneling-like
effects in the metastable Higgs potential. Here, since $\lambda^{\prime}%
\sim\Omega_{\Lambda}>0$ is positive, the renewal in regions $\left\vert
\tau\right\vert >T_{+}/2$ has no relation to the possible tunneling of the
Higgs in a metastable renormalized Higgs potential. A physical interpretation
of the reason behind this cyclic-universe type periodicity remains open.}.

The spiking behavior of $\vec{s}\left(  \pm T_{0}\right)  =\pm\infty$ when
$z\left(  \pm T_{0}\right)  =0$ at the cosmological singularities, as seen in
Figs.5 is quite general$.$ This occurs for all values of the integration
constants $\left(  \vec{s}_{0},\vec{p}\right)  ,$ so the simultaneous
divergence of the scalar $\sigma\left(  \tau\right)  $ and anisotropy $\left(
\alpha_{1}\left(  \tau\right)  ,\alpha_{2}\left(  \tau\right)  \right)  $ when
$z\left(  \tau\right)  $ hits zero cannot be avoided. The constant $\vec
{s}_{0}$ is represented in Fig.5 as the point the vertical axis intersects the
curve $\vec{s}\left(  \tau\right)  $. To emphasize the spiking property of
$\vec{s}\left(  \tau\right)  $, I also produce a parametric plot $\left(
z\left(  \tau\right)  ,\vec{s}\left(  \tau\right)  \right)  $ that amounts to
a plot of $\vec{s}$ as a function of $z,$ or vice-versa, as seen in Fig.6.

It is revealing to reorganize the four degrees of freedom $\left(  z,\vec
{s}\right)  $ into the form $\left(  \phi_{\hat{p}},h_{\hat{p}}\right)  $
defined as follows
\begin{equation}%
\begin{array}
[c]{c}%
u_{\hat{p}}\equiv\sqrt{\left\vert z\right\vert }e^{\hat{p}\cdot\vec{s}%
},\;v_{\hat{p}}\equiv\sqrt{\left\vert z\right\vert }e^{-\hat{p}\cdot\vec{s}}\\
\phi_{\hat{p}}\left(  \tau\right)  =\frac{1}{2}\left(  u_{\hat{p}}\left(
\tau\right)  +\varepsilon_{z}v_{\hat{p}}\left(  \tau\right)  \right)
,\;h_{\hat{p}}\left(  \tau\right)  =\frac{1}{2}\left(  u_{\hat{p}}\left(
\tau\right)  -\varepsilon_{z}v_{\hat{p}}\left(  \tau\right)  \right)  ,
\end{array}
\label{UandV}%
\end{equation}
where $\hat{p}=\vec{p}/\left\vert \vec{p}\right\vert $ is a unit vector. These
generalize $\left(  \phi,h\right)  $ by including anisotropy through the
angles $\hat{p}.$ A parametric plot of these functions is given in Fig.7,
where the trajectory along the arrows show the progress as $\tau$ increases
from $-T_{+}/2$ to $T_{+}/2.$ The motion is in the plane slicing the touching
cones and containing the vector $\hat{p}$ perpendicular to the vertical axis.
A revolution of this figure around the vertical axis is equivalent to changing
the direction of the vector $\hat{p}.$ The interior of the cones correspond to
the gravity regions II \& IV while the exterior of the cones correspond to the
antigravity regions I \& III. The trajectory of the universe in this figure is
equivalent to the solution for $\left(  z\left(  \tau\right)  ,\vec{s}\left(
\tau\right)  \right)  $ plotted in Fig.6 but now re-plotted in these new
coordinates. The figure visually shows that the universe contracts in region
IV, passes in all directions $\hat{p}$ through a 4-dimensional pin hole,
$\phi_{\hat{p}}\left(  \tau_{c}\right)  =h_{\hat{p}}\left(  \tau_{c}\right)
=0$ at the crunch time $\tau_{c}=-T_{0},$ expands a little into antigravity,
and turns around during antigravity to pass again through a pin hole,
$\phi_{\hat{p}}\left(  \tau_{b}\right)  =h_{\hat{p}}\left(  \tau_{b}\right)
=0$ at the bang time $\tau_{b}=+T_{0}$, and then expands in region II.
\textit{ }%
\begin{center}
\includegraphics[
height=2.6521in,
width=2.4141in
]%
{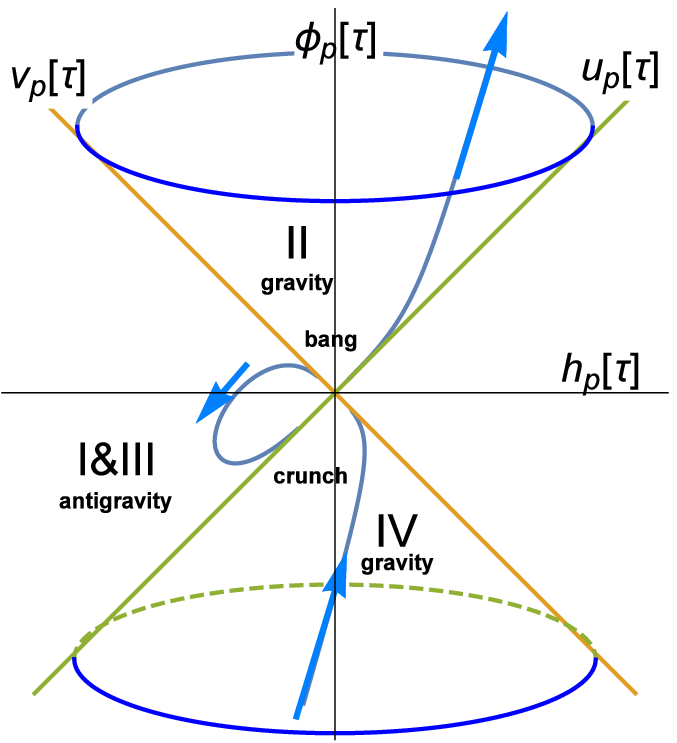}%
\\
Fig.(7) - {\protect\small Evolution of the universe must pass through all
gravity and antigravity regions but only by shrinking to zero size,
IV}$\rightarrow0\rightarrow${\protect\small (I\&II)}$\rightarrow0\rightarrow
${\protect\small II. At zero size }$\phi_{\hat{p}}=h_{\hat{p}}=0~$
{\protect\small  and their ratio is }$\pm1~${\protect\small  as seen by the
tangential slopes to the horizons }$u_{\hat{p}}=0~$ {\protect\small  or
}$v_{\hat{p}}=0.$%
\end{center}

This plot is a generalization of a similar one in \cite{crunchBang}, but shows
more dramatically the pin hole effect in four dimensions including anisotropy.
This emphasizes the theorem stated below that the scalar and anisotropy
degrees of freedom spike to infinity at the singularity. In the re-organized
pin hole version this makes it evident that the dynamics is one of an
unavoidable attractor to the tips of the cones. The trajectory of the universe
cannot cross the cosmological horizons represented by the walls of the cones
in Fig.7. The trajectory can evolve to the neighboring Rindler regions only by
going through the \textquotedblleft pin hole\textquotedblright\ and
tangentially to the walls of the cones. This result is a remarkable unique
cosmological prediction for the initial conditions of the universe at the
\textquotedblleft beginning\textquotedblright. This is a sufficiently
important cosmological prediction of the EFTC introduced in section
\ref{theory}, that I will highlight it as a theorem:

\textit{Theorem: The attractor mechanism displayed in Fig.7 produces
dynamically unique initial values for all degrees of freedom at cosmological
singularities }$z=0$\textit{. Quantitatively, at both the bang/crunch, the
fields }$\left(  \phi_{\hat{p}}\left(  \tau\right)  ,h_{\hat{p}}\left(
\tau\right)  \right)  $\textit{ must vanish simultaneously for every \^{p}
while their ratio must be plus or minus one, }%
\begin{equation}
\phi_{\hat{p}}\left(  \tau_{b/c}\right)  =h_{\hat{p}}\left(  \tau
_{b/c}\right)  =0,\;\frac{h_{\hat{p}}\left(  \tau_{b/c}\right)  }{\phi
_{\hat{p}}\left(  \tau_{b/c}\right)  }=\pm1. \label{pinhole}%
\end{equation}
\textit{Thus, the universe must pass through a \textquotedblleft pin
hole\textquotedblright\ while matching all degrees of freedom on both sides of
the pin hole in all directions }$\hat{p}$\textit{. An alternative description
of the \textquotedblleft pin hole\textquotedblright\ initial values in the 4D
variables, }$\left(  z\left(  \tau_{b/c}\right)  ,\vec{s}\left(  \tau
_{b/c}\right)  \right)  =\left(  0,\pm\vec{\infty}\right)  ,$\textit{ is given
in Fig.6.} \textit{This theorem emerges not only in the classical analysis
given above, but also in the quantum analysis given in section \ref{QM2}. }

A remark about Misner's \textquotedblleft
mixmaster\ universe\textquotedblright\ problem \cite{misner} is in order. The
classical solution for $\left(  z\left(  \tau\right)  ,\vec{s}\left(
\tau\right)  \right)  $ that includes anisotropy, as given in (\ref{zLambda}%
,\ref{stau}) and plotted in Figs.(4-7), clearly did not suffer from
this problem even in the case of Bianchi-IX metric $\left(
k=1\right)  $, so it invites an explanation. This solution was
obtained under the assumption that $V_{K}\left(
\alpha_{1},\alpha_{2}\right)  $ was replaced by a constant as in
(\ref{Vapprox}), so naturally the rattling around that causes
Misner's mixmaster did not happen in the constant potential.
Therefore one must re-examine if the approximation of neglecting the
details of the potential is valid. I have performed the following
computation: I inserted the solution above into the full action
(\ref{Sminifh}) without approximating $V_{K}\left(
\alpha_{1},\alpha_{2}\right)  $ and asked under what conditions the
potential energy \textit{for this generic solution as a function of
}$\tau$ can be neglected as compared to its kinetic energy as a
function of $\tau$ while satisfying $\mathcal{H}=0$. The answer to
this question is that, it is true that the kinetic terms \textit{for
the solution} do dominate and the potential term $z\left(
\tau\right)  V_{K}\left(  \alpha_{1}\left( \tau\right)
,\alpha_{2}\left(  \tau\right)  \right)  $ vanishes as
$\tau\rightarrow\tau_{b/c}$. In more detail, based on this analysis,
the solutions (\ref{zLambda},\ref{stau}) are legitimate
approximations as long as
the conserved momenta satisfy the inequality, $0<4\left\vert p_{1,2}%
\right\vert \lesssim\left\vert p_{3}\right\vert \lesssim\left\vert \vec
{p}\right\vert $. It is significant that without a non-zero $\left\vert
p_{3}\right\vert $ it is not possible to satisfy this inequality, which means
the Higgs is crucial. This simple result for the avoidance of mixmaster is
consistent with the considerably more complicated BKL analysis \cite{BKL} that
concludes the mixmaster is avoidable when there are scalars in the theory.

In the current EFTC the presence of the Higgs is essential for avoiding the
mixmaster, as well as for creating the dynamical attractor that predicts
initial conditions. So the Higgs in this EFTC has important cosmological roles
in shaping our universe.

\section{Quantum wavefunction - 2 \label{QM2}}

I now turn to the solution of the WdWe (\ref{WdWezs}). First I analyze its
full form before the suggested approximation strategy in Eq.(\ref{Vapprox}%
-\ref{gaussian}) and confront the problem of continuity in general. The
partial differential equation (\ref{WdWfh}) or (\ref{WdWezs}), in any of the
$\left(  \phi,h\right)  $ or $\left(  u,v\right)  $ or $\left(  z,\sigma
\right)  $ bases, is a mathematically well-defined quantum problem that
includes continuous passage through singularities at $u=\left(  \phi+h\right)
=0$ or $v=\left(  \phi-h\right)  =0$ (which means E-frame singularity at $z=0$
or $a_{E}=0$). Continuity of the wavefunction through the global coordinates
at $u=0$ or $v=0$ is not trivial and poses a main challenge whose full
resolution is given below in Eq.(\ref{WaveUniv}). The tricky technical problem
is outlined as follows. The total potential in (\ref{WdWfh},\ref{WdWezs}) is
dominated by the $\Omega_{c}$-term close to the singularity $z\rightarrow0.$
Since the potentials become immaterial in this limit, one may go to momentum
space and solve the two basic solutions of (\ref{WdWfh}) in $\left(
u,v\right)  $ basis, ($4\partial_{u}\partial_{v}-\frac{1}{uv}\left(
\partial_{1}^{2}+\partial_{2}^{2}\right)  +\Omega_{c}+\cdots)\Psi=0,$ near the
singularity $z=uv\sim0.$ These are similar to (\ref{packet})
\begin{equation}
\Psi:A_{\pm}\left(  \vec{p}\right)  e^{-ip_{1}\alpha_{1}-ip_{2}\alpha_{2}%
}u^{-i\left(  p_{3}\pm\left\vert \vec{p}\right\vert \right)  /2}v^{i\left(
p_{3}\mp\left\vert \vec{p}\right\vert \right)  /2}S_{\pm\left\vert \vec
{p}\right\vert }\left(  uv\right)  , \label{nearSing}%
\end{equation}
with the leading terms in $S_{\pm\left\vert \vec{p}\right\vert }\left(
uv\right)  \simeq\left(  1+O\left(  \Omega_{c}uv\right)  \right)  ,$ and any
real $\vec{p}\equiv\left(  p_{1},p_{2},p_{3}\right)  .$ The complete solution
$\Psi$ near $z=uv=0$ is the general linear superposition over these two basic
solutions summed up as in (\ref{packet}), although now this is without
approximating the potentials, but examining only the vicinity of $z=0.$ So,
the form (\ref{nearSing}) near the singularity persists despite the potentials.

The problem with continuity is that, at either $u\rightarrow0$ when $v$ is
finite, or at $v\rightarrow0$ when $u$ is finite, i.e. at any point on the
horizons in Fig.1, the basic solutions (\ref{nearSing}) seem to oscillate
wildly so that their values on either side of the horizons appear to be
undeterminable. However, under the integral $\int d^{3}p$, with sufficiently
smooth $A_{\pm}\left(  \vec{p}\right)  ,$ such wildly oscillating integrands
produce a definite vanishing value for the integral, thus giving
$\Psi\rightarrow0$ at the horizons. If this were true, the coefficients
$A_{\pm}\left(  \vec{p}\right)  $ would be chosen independently on either side
of each horizon in Fig.1, so that a solution $\Psi$ within each of the four
quadrants would vanish at the horizons, and be independent from the solutions
within all other quadrants in Fig.1.

A discontinuous wavefunction due to uncontrollable oscillations is rejected
since it is not a solution of \ref{WdWfh} and furthermore gives problems with
the Hermiticity of the Hamiltonian just as in the case of the singular
$\left(  -1/r^{2}\right)  $ potential in ordinary quantum mechanics. Until now
this remained an unsettled problem\footnote{The connection to $\left(
-1/r^{2}\right)  $ becomes evident in Eq.(\ref{WdWSch}) that displays the
attractive $\left(  -1/z^{2}\right)  $ potential as the leading singular term
of the potential, corresponding precisely to the cosmological singularity of
order $a_{E}^{-6}$ in the Friedmann equation (\ref{Omegas}). The unique
solution to this problem in the setting of this paper, using the complete set
of normalized and continuous wavepackets, given in the next paragraph was not
attempted before.} that goes back to Von Neumann. Various suggested solutions
are non-unique and differ physically in different physical approximations to a
regulated $\left(  -1/r^{2}\right)  $ \cite{oneoverz2}. A resolution seems to
be in sight by using wavepacket solutions as just outlined above, because the
wild oscillations are controlled in a complete set of normalizable wavepackets
and continuity appears to be satisfied with $\Psi\rightarrow0$ at both sides
of all horizons. However, this is problematic for a geodesically complete
cosmology advanced in this paper, because there seems to be no relation
between the wavefunctions for the past (region IV), the future (region II) or
antigravity (regions I \& III), and therefore no information from the past
seems to survive to the future of the big bang.

The very tricky subtlety that resolves this problem is that the story for
vanishing wavepackets $\Psi\rightarrow0$ at all horizons is not true. This is
because, according to (\ref{coordTransf}) one can rewrite the basic solutions
near the singularity (\ref{nearSing}) as follows
\begin{equation}
e^{-ip_{1}\alpha_{1}-ip_{2}\alpha_{2}}u^{-i\left(  p_{3}\pm\left\vert \vec
{p}\right\vert \right)  /2}v^{i\left(  p_{3}\mp\left\vert \vec{p}\right\vert
\right)  /2}\sim e^{-i\vec{p}\cdot\vec{s}}\sqrt{\left\vert z\right\vert }^{\mp
i\left\vert \vec{p}\right\vert }=\left(  \sqrt{\left\vert z\right\vert }%
e^{\pm\hat{p}\cdot\vec{s}}\right)  ^{\mp i\left\vert \vec{p}\right\vert },
\label{cuts1}%
\end{equation}
where $\hat{p}=\vec{p}/\left\vert \vec{p}\right\vert .$ Then, when $\left\vert
z\right\vert \rightarrow0,$ there are regions of the integral $\int d^{3}p$
that do not oscillate wildly by having $\left\vert \hat{p}\cdot\vec
{s}\right\vert \rightarrow\infty$ in tandem with $\sqrt{\left\vert
z\right\vert }\rightarrow0$ so that either $\sqrt{\left\vert z\right\vert
}e^{\hat{p}\cdot\vec{s}}$ or $\sqrt{\left\vert z\right\vert }e^{-\hat{p}%
\cdot\vec{s}}$ remains finite while the other goes to zero. Defining
$u_{\hat{p}}\equiv\sqrt{\left\vert z\right\vert }e^{\hat{p}\cdot\vec{s}%
},\;\;v_{\hat{p}}\equiv\sqrt{\left\vert z\right\vert }e^{-\hat{p}\cdot\vec{s}%
},$ as in (\ref{UandV}), the two solutions near $z=0$ in (\ref{cuts1}) take
the form
\begin{equation}
u_{\hat{p}}^{-i\left\vert \vec{p}\right\vert }~~\text{or~~}v_{\hat{p}%
}^{i\left\vert \vec{p}\right\vert }, \label{cuts2}%
\end{equation}
such that $u_{\hat{p}}^{-i\left\vert \vec{p}\right\vert }$ remains finite when
$v_{\hat{p}}^{i\left\vert \vec{p}\right\vert }$ oscillates wildly and vice
versa. So there is a part of the superposition in $\Psi$ at each horizon in
which the wild oscillations do not occur, thus making a finite contribution to
$\Psi.$ This region is always in the mini-superspace region where $\left\vert
\hat{p}\cdot\vec{s}\right\vert \rightarrow+\infty$ as $z\rightarrow0$ for all
available directions $\hat{p}.$ This finite part of $\Psi$ has to be the same
on either side of each horizon in order to have a continuous wavefunction
between neighboring quadrants in the full 4D mini-superspace $\left(
z,\vec{s}\right)  $. This then resolves the problem of continuity which is now
achieved simply by analytic continuation in the complex $u_{\hat{p}}%
,v_{\hat{p}}$ planes. The details of this continuity mechanism was discussed
recently in great detail in \cite{barsAraya-rindler} for the case $\hat
{p}=\left(  0,0,\pm1\right)  .$ The techniques are the same and it amounts to
a simple generalization of \cite{barsAraya-rindler} to the general $\hat{p}.$

The outcome is that, close to the singularity $z=0$ the probability amplitude
$\Psi$ gets its support from the region where the scalar and anisotropy
degrees of freedom $\vec{s}$ diverge, in agreement with Fig.6. Equivalently,
the support for non-zero $\Psi$ close to the singularity is found in the
equivalent variables $\left(  u_{\hat{p}},v_{\hat{p}}\right)  $ in the
neighborhood of the pin hole as in Fig.7. Hence the \textit{theorem for unique
initial conditions} emerges both in classical and quantum dynamics.

With this information one can now carry out the task of imposing continuity of
the wavefunction as in \cite{barsAraya-rindler}, and find that the past,
antigravity and future wavepacket coefficients are related to each other. This
leads to the complete general analytic solution of the full continuous
wavefunction $\Psi$, with \textit{only one set of independent wavepacket
coefficients} $\left(  a\left(  \vec{p}\right)  ,b^{\dagger}\left(  \vec
{p}\right)  \right)  $ that appear in the solutions in the various quadrants,
as displayed in Eq.(\ref{WaveUniv}).
\begin{equation}%
\begin{array}
[c]{l}%
\Psi_{II}\left(  z,\vec{s}\right)  \overset{z>0}{=}\int_{-\infty}^{\infty
}d^{3}p\left[  a\left(  \vec{p}\right)  \frac{e^{-i\vec{p}\cdot\vec{s}}%
\sqrt{z}^{-i\left\vert \vec{p}\right\vert }}{\sqrt{16\pi^{3}\left\vert \vec
{p}\right\vert }}S_{\left\vert \vec{p}\right\vert }\left(  z\right)
+b^{\dagger}\left(  \vec{p}\right)  \frac{e^{i\vec{p}\cdot\vec{s}}\sqrt
{z}^{i\left\vert \vec{p}\right\vert }}{\sqrt{16\pi^{3}\pi\left\vert \vec
{p}\right\vert }}S_{-\left\vert \vec{p}\right\vert }\left(  z\right)  \right]
,\\
\Psi_{I\&III}\left(  z,\vec{s}\right)  \overset{z<0}{=}\int_{-\infty}^{\infty
}d^{3}p\left[
\begin{array}
[c]{c}%
a\left(  \vec{p}\right)  \left\{  \frac{e^{-i\vec{p}\cdot\vec{s}}\sqrt
{-z}^{-i\left\vert \vec{p}\right\vert }S_{\left\vert \vec{p}\right\vert
}\left(  z\right)  }{\sqrt{16\pi^{3}\left\vert \vec{p}\right\vert }}%
+\frac{e^{-i\vec{p}\cdot\vec{s}}\sqrt{-z}^{i\left\vert \vec{p}\right\vert
}S_{-\left\vert \vec{p}\right\vert }\left(  z\right)  }{\sqrt{16\pi
^{3}\left\vert \vec{p}\right\vert }}\frac{\left(  \frac{\Omega_{\Lambda}}%
{18}\right)  ^{i\frac{\left\vert \vec{p}\right\vert }{6}}\Gamma\left(
-i\frac{\left\vert \vec{p}\right\vert }{3}\right)  }{\Gamma\left(
i\frac{\left\vert \vec{p}\right\vert }{3}\right)  }\right\} \\
+b^{\dagger}\left(  \vec{p}\right)  \left\{  \frac{e^{i\vec{p}\cdot\vec{s}%
}\sqrt{-z}^{i\left\vert \vec{p}\right\vert }S_{-\left\vert \vec{p}\right\vert
}\left(  z\right)  }{\sqrt{16\pi^{3}\pi\left\vert \vec{p}\right\vert }}%
+\frac{e^{i\vec{p}\cdot\vec{s}}\sqrt{-z}^{-i\left\vert \vec{p}\right\vert
}S_{\left\vert \vec{p}\right\vert }\left(  z\right)  }{\sqrt{16\pi
^{3}\left\vert \vec{p}\right\vert }}\frac{\left(  \frac{\Omega_{\Lambda}}%
{18}\right)  ^{-i\frac{\left\vert \vec{p}\right\vert }{6}}\Gamma\left(
i\frac{\left\vert \vec{p}\right\vert }{3}\right)  }{\Gamma\left(
-i\frac{\left\vert \vec{p}\right\vert }{3}\right)  }\right\}
\end{array}
\right] \\
\Psi_{IV}\left(  z,\vec{s}\right)  \overset{z>0}{=}\int_{-\infty}^{\infty
}d^{3}p\left[  a\left(  \vec{p}\right)  \frac{e^{-i\vec{p}\cdot\vec{s}}%
\sqrt{z}^{i\left\vert \vec{p}\right\vert }S_{-\left\vert \vec{p}\right\vert
}\left(  z\right)  }{\sqrt{16\pi^{3}\left\vert \vec{p}\right\vert }}%
\frac{\left(  \frac{\Omega_{\Lambda}}{18}\right)  ^{i\frac{\left\vert \vec
{p}\right\vert }{6}}\Gamma\left(  -i\frac{\left\vert \vec{p}\right\vert }%
{3}\right)  }{\Gamma\left(  i\frac{\left\vert \vec{p}\right\vert }{3}\right)
}+b^{\dagger}\left(  \vec{p}\right)  \frac{e^{i\vec{p}\cdot\vec{s}}\sqrt
{z}^{-i\left\vert \vec{p}\right\vert }S_{\left\vert \vec{p}\right\vert
}\left(  z\right)  }{\sqrt{16\pi^{3}\left\vert \vec{p}\right\vert }}%
\frac{\left(  \frac{\Omega_{\Lambda}}{18}\right)  ^{-i\frac{\left\vert \vec
{p}\right\vert }{6}}\Gamma\left(  i\frac{\left\vert \vec{p}\right\vert }%
{3}\right)  }{\Gamma\left(  -i\frac{\left\vert \vec{p}\right\vert }{3}\right)
}\right]  .
\end{array}
\label{WaveUniv}%
\end{equation}

The form of (\ref{WaveUniv}) is consistent with the form (\ref{nearSing}) if
taken only near $z\sim0$ without approximating the potentials. It is also
consistent with the form (\ref{packet}) if the suggested approximation
strategy of section \ref{QM1} is applied. In the latter case the functions
$S_{\pm\left\vert \vec{p}\right\vert }\left(  z\right)  $ that appear in these
expressions is computed analytically and given below; in that case the
solution for the wavefunction above is valid in all regions of the
geodesically complete superspace depicted in Fig.7. The labels $II,I\&III,IV$
on the wavefunctions in (\ref{WaveUniv}) refer to the corresponding regions in
Fig.7, namely the inside and outside regions of the cones. As expected from
the qualitative discussion of Figs.2,3, the $\Psi_{II}\left(  z,\vec
{s}\right)  $ and $\Psi_{IV}\left(  z,\vec{s}\right)  ,$ that are the future
and past positive-$z$ gravity sectors respectively, are oscillatory
(scattering-type wave packets), while $\Psi_{I\&III}\left(  z,\vec{s}\right)
$ that are at negative-$z$ have fast decaying asymptotic behavior as
$z\rightarrow-\infty$ deep into the antigravity region. The asymptotic
behavior of the functions $S_{\pm\left\vert \vec{p}\right\vert }\left(
z\right)  $ as $z\rightarrow\pm\infty$ were used in order to fix all relative
coefficients in these expressions so that the correct physical behavior is
obtained (oscillatory in II \& IV and decay in I \& III). So, the probability
of the universe to spend time in the antigravity regions I\&III during its
evolution is limited by the fast decay of $\Psi_{I\&III}\left(  z,\vec
{s}\right)  $ away from the $z=0$ singularity. The simple explanation for this
behavior is easily understood from the shape of the potential barrier as
discussed above in relation to Figs.2,3.

How continuity works requires some guidance. Continuity at the future horizons
in Fig.7 requires the wavefunction $\Psi_{I\&III}\left(  z,\vec{s}\right)  $
and $\Psi_{II}\left(  z,\vec{s}\right)  $ to share the same $\left(  a\left(
\vec{p}\right)  ,b^{\dagger}\left(  \vec{p}\right)  \right)  $ coefficients as
follows. The part proportional to $a\left(  \vec{p}\right)  $ in $\Psi
_{II}\left(  z,\vec{s}\right)  $ at $z>0$ is directly related to \textit{the
first term} in $\Psi_{I\&III}\left(  z,\vec{s}\right)  $ at $z<0,$ as written
in (\ref{WaveUniv}). The second term proportional to $a\left(  \vec{p}\right)
$ in $\Psi_{I\&III}\left(  z,\vec{s}\right)  $ vanishes at the future horizon
due to the fast oscillations. The part proportional to $b^{\dagger}\left(
\vec{p}\right)  $ in comparing $\Psi_{I\&III}\left(  z,\vec{s}\right)  $ and
$\Psi_{II}\left(  z,\vec{s}\right)  $ at the future horizons in Fig.7 works
exactly the same way. For the past horizon, the arguments are quite similar,
and in this way the \textit{second terms} in $\Psi_{I\&III}\left(  z,\vec
{s}\right)  $ proportional $\left(  a\left(  \vec{p}\right)  ,b^{\dagger
}\left(  \vec{p}\right)  \right)  $ are analytically continued from $z<0$ to
$z>0$ thus connecting to $\Psi_{II}\left(  z,\vec{s}\right)  $ as written in
(\ref{WaveUniv}), while the first terms in $\Psi_{I\&III}\left(  z,\vec
{s}\right)  $ proportional to $\left(  a\left(  \vec{p}\right)  ,b^{\dagger
}\left(  \vec{p}\right)  \right)  $ vanish at $z=0$ for the past horizon due
to the fast oscillations. This continuity of the overall $\Psi_{II,I\&III,IV}$
can be rephrased in terms of the $\left(  u_{\hat{p}},v_{\hat{p}}\right)  $
basis, as genuine analytic continuation around cuts in the complex planes of
the variables $\left(  u_{\hat{p}},v_{\hat{p}}\right)  $ in every direction
$\hat{p}.$ This is explained in great detail in \cite{barsAraya-rindler} for
the special direction $\hat{p}=\left(  0,0,\pm1\right)  .$

The form (\ref{WaveUniv}) of the general solution is complete because it is
based on the complete set of solutions for the basis functions and can be
applied to all possible physical circumstances. Like in field theory, the
wavepacket coefficients $\left(  a\left(  \vec{p}\right)  ,b^{\dagger}\left(
\vec{p}\right)  \right)  $ can be interpreted as analogs of
creation/annihilation operators for positive/negative frequency massless plane
waves $e^{-i\vec{p}\cdot\vec{s}\pm i\left\vert \vec{p}\right\vert \ln
\sqrt{\left\vert z\right\vert }}$ at the horizons, since $S_{\left\vert
\vec{p}\right\vert }\left(  z\right)  \overset{z\rightarrow0}{\rightarrow}1,$
while interpreting $\ln\sqrt{\left\vert z\right\vert }$ as the analog of time
close to the horizons in Fig.7. This is similar to the case of the plane wave
basis at black hole horizons. One may set up a Bogoliubov transformation
between these and creation/annihilation operators for plane waves at the
asymptotics of region II, similar to the case of black holes or similar to the
one given in \cite{barsAraya-rindler} for the special direction $\hat
{p}=\left(  0,0,\pm1\right)  .$

For the physical application of interest in this paper, namely the
wavefunction for the universe, additional boundary conditions are required in
the asymptotic regions of II and IV. Namely, the wavepacket $\Psi_{IV}\left(
z,\vec{s}\right)  $ must have \textit{only incoming asymptotic waves} (not
horizon waves), where incoming is defined in region IV at $z\rightarrow
+\infty$ as the leading oscillatory behavior $e^{i\omega z^{3/2}}$ with
positive $\omega$. Once this is imposed in momentum space on $\Psi_{IV}\left(
z,\vec{s}\right)  $ it turns out the wavepacket $\Psi_{II}\left(  z,\vec
{s}\right)  $ automatically has only outgoing asymptotic waves because of the
relations of the wavepacket coefficients in the different regions as displayed
in (\ref{WaveUniv}). This asymptotic boundary condition determines
$b^{\dagger}\left(  -\vec{p}\right)  $ as a function of $a\left(  \vec
{p}\right)  $
\begin{equation}
b^{\dagger}\left(  -\vec{p}\right)  =a\left(  \vec{p}\right)  \frac
{\Gamma\left(  -i\frac{\left\vert \vec{p}\right\vert }{3}\right)  }%
{\Gamma\left(  i\frac{\left\vert \vec{p}\right\vert }{3}\right)  }\left(
\frac{\Omega_{\Lambda}}{18}\right)  ^{i\frac{\left\vert \vec{p}\right\vert
}{3}}e^{-\frac{\pi\left\vert \vec{p}\right\vert }{3}}. \label{BasA}%
\end{equation}
To get this result, the asymptotic form of $S_{\left\vert \vec{p}\right\vert
}\left(  z\right)  $ given in Eq.(\ref{case1}) below is used.

In addition, $\left(  a\left(  \vec{p}\right)  ,b^{\dagger}\left(  \vec
{p}\right)  \right)  $ must be limited to Gaussians as explained in
(\ref{gaussian}) as part of the approximation strategy in section (\ref{QM1}),
therefore%
\begin{equation}
a\left(  \vec{p}\right)  =Ae^{-\left(  p_{1}^{2}+p_{2}^{2}\right)
/2\Omega_{\alpha}}e^{-p_{3}^{2}/2\Omega_{\sigma}}, \label{a(p)}%
\end{equation}
and similarly for $b^{\dagger}\left(  \vec{p}\right)  $ related by
(\ref{BasA}). The wavefunction in this form has no remaining unknown
parameters since the overall $A$ is just a normalization factor.

Major conclusions of this discussion are:

\begin{enumerate}
\item Anisotropy and Higgs $\left\vert \hat{p}\cdot\vec{s}\right\vert $ must
be at infinity for all available directions $\hat{p}$ when $a_{E}=0,$ or
equivalently at the pin hole of Fig.7 in terms of the $\left(  u_{\hat{p}%
},v_{\hat{p}}\right)  $ basis.

\item The general solution for the wavefunction for all physical applications
as given in (\ref{WaveUniv}) depends only on one set of wavepacket
coefficients $\left(  a\left(  \vec{p}\right)  ,b^{\dagger}\left(  \vec
{p}\right)  \right)  ,$ analogous to the creation/annihilation operators of a
field, that are shared in all patches of the geodesically complete
mini-superspace. From this it is evident that, if the $\left(  a\left(
\vec{p}\right)  ,b^{\dagger}\left(  \vec{p}\right)  \right)  $ were quantized,
the resultant Fock space would be complete, and would describe the physics for
all gravity and antigravity patches, in a \textit{unitary Hilbert space}. This
is in agreement with the complementary discussion given in \cite{barsJames} on
unitarity and stability.

\item The wavefunction for the universe is given by the restricted form of
$\left(  a\left(  \vec{p}\right)  ,b^{\dagger}\left(  \vec{p}\right)  \right)
$ in Eqs.(\ref{BasA},\ref{a(p)}). In this completely fixed form the
wavefunction has a unique dynamically generated initial value at the
\textquotedblleft beginning\textquotedblright. This predicted last form is a
topic of discussion in section \ref{QM3}.
\end{enumerate}

\section{Quantum wavefunction - 3 \label{QM3}}

The wavefunction for the universe derived in the previous section is
simplified as follows. It has no unspecified attributes and depends only on
the phenomenologically measured parameters $\left(  \Omega_{\Lambda}%
,\Omega_{c},\Omega_{K},\Omega_{\sigma},\Omega_{\alpha}\right)  $ that appear
in the Friedmann equation (\ref{Omegas})%
\begin{equation}%
\begin{array}
[c]{l}%
\Psi_{II}\left(  z,\vec{s}\right)  \overset{z>0}{=}A\int_{-\infty}^{\infty
}\frac{d^{3}p}{\sqrt{\left(  2\pi\right)  ^{3}2\left\vert \vec{p}\right\vert
}}~e^{-\frac{p_{3}^{2}}{2\Omega_{\sigma}}}e^{-\frac{p_{1}^{2}+p_{2}^{2}%
}{2\Omega_{\alpha}}}e^{-i\vec{p}\cdot\vec{s}}H_{\left\vert \vec{p}\right\vert
}^{+}\left(  z\right) \\
\Psi_{I\&III}\left(  z,\vec{s}\right)  \overset{z<0}{=}A\int_{-\infty}%
^{\infty}\frac{d^{3}p}{\sqrt{\left(  2\pi\right)  ^{3}2\left\vert \vec
{p}\right\vert }}~e^{-\frac{p_{3}^{2}}{2\Omega_{\sigma}}}e^{-\frac{p_{1}%
^{2}+p_{2}^{2}}{2\Omega_{\alpha}}}e^{-i\vec{p}\cdot\vec{s}}\left(
H_{\left\vert \vec{p}\right\vert }^{+}\left(  z\right)  +H_{\left\vert \vec
{p}\right\vert }^{-}\left(  z\right)  \right) \\
\Psi_{IV}\left(  z,\vec{s}\right)  \overset{z>0}{=}A\int_{-\infty}^{\infty
}\frac{d^{3}p}{\sqrt{\left(  2\pi\right)  ^{3}2\left\vert \vec{p}\right\vert
}}~e^{-\frac{p_{3}^{2}}{2\Omega_{\sigma}}}e^{-\frac{p_{1}^{2}+p_{2}^{2}%
}{2\Omega_{\alpha}}}e^{-i\vec{p}\cdot\vec{s}}H_{\left\vert \vec{p}\right\vert
}^{-}\left(  z\right)
\end{array}
\label{WaveUniverse}%
\end{equation}
where the overall $A$ is fixed by normalization, and the functions
$H_{\left\vert \vec{p}\right\vert }^{\pm}\left(  z\right)  $ are
\begin{equation}%
\begin{array}
[c]{l}%
H_{\left\vert \vec{p}\right\vert }^{+}\left(  z\right)  =\left(
\sqrt{\left\vert z\right\vert }^{-i\left\vert \vec{p}\right\vert
}S_{\left\vert \vec{p}\right\vert }\left(  z\right)  +\sqrt{\left\vert
z\right\vert }^{i\left\vert \vec{p}\right\vert }S_{-\left\vert \vec
{p}\right\vert }\left(  z\right)  ~e^{-\frac{\pi\left\vert \vec{p}\right\vert
}{3}}\frac{\left(  \frac{\Omega_{\Lambda}}{18}\right)  ^{i\frac{\left\vert
\vec{p}\right\vert }{3}}\Gamma\left(  -i\frac{\left\vert \vec{p}\right\vert
}{3}\right)  }{\Gamma\left(  i\frac{\left\vert \vec{p}\right\vert }{2}\right)
}\right) \\
H_{\left\vert \vec{p}\right\vert }^{-}\left(  z\right)  =\left(  \sqrt
{z}^{i\left\vert \vec{p}\right\vert }S_{-\left\vert \vec{p}\right\vert
}\left(  z\right)  \frac{\Gamma\left(  -i\frac{\left\vert \vec{p}\right\vert
}{3}\right)  }{\Gamma\left(  i\frac{\left\vert \vec{p}\right\vert }{3}\right)
}+\sqrt{z}^{-i\left\vert \vec{p}\right\vert }S_{\left\vert \vec{p}\right\vert
}\left(  z\right)  ~e^{-\frac{\pi\left\vert \vec{p}\right\vert }{3}}\right)
\left(  \frac{\Omega_{\Lambda}}{18}\right)  ^{i\frac{\left\vert \vec
{p}\right\vert }{6}}%
\end{array}
\label{H+-}%
\end{equation}
Finally there remains to give an explicit $S_{\pm\left\vert \vec{p}\right\vert
}\left(  z\right)  $ that depends on $\left(  \left\vert \vec{p}\right\vert
,\Omega_{\Lambda},\Omega_{c},\Omega_{K}\right)  .$ The $S_{\pm\left\vert
\vec{p}\right\vert }\left(  z\right)  $ that solves the WdWe (\ref{WdWSch})
for the potential $V$ in Fig.2 with all parameters $\left(  \left\vert \vec
{p}\right\vert ,\Omega_{\Lambda},\Omega_{c},\Omega_{K}\right)  $ non-zero is
not known analytically at this time, although I think this could be obtained.
It can certainly be determined numerically or other approximations, such as
WKB. However, I have constructed analytic solutions for all the cases listed
below in which some of these parameters $\left(  \left\vert \vec{p}\right\vert
,\Omega_{\Lambda},\Omega_{c},\Omega_{K}\right)  $ are set to zero.

The most useful approximation is case-1 given below. This case captures best
the physical features of the full $S_{\pm\left\vert \vec{p}\right\vert
}\left(  z\right)  $ because the potential $\tilde{V}\left(  z\right)  $ shown
in Fig.3 agrees with the leading terms of the full $V\left(  z\right)  $ at
both limits $z\rightarrow0^{\pm}$ and $z\rightarrow\pm\infty$
\begin{equation}%
\begin{array}
[c]{l}%
\text{Case-1:\ }\left(  \Omega_{c},\Omega_{K}\right)  \rightarrow0,\;V\left(
z\right)  \rightarrow\tilde{V}\left(  z\right)  =-\left(  \frac{\vec{p}^{2}%
+1}{4z^{2}}+\frac{\Omega_{\Lambda}}{2}z\right)  ,\;\\
\mathcal{H}\Psi=\left(  \partial_{\phi}^{2}-\partial_{h}^{2}-\frac
{\partial_{1}^{2}+\partial_{2}^{2}}{\phi^{2}-h^{2}}+2\Omega_{\Lambda}\left(
\phi^{2}-h^{2}\right)  ^{2}\right)  \Psi=0,\\
S_{\left\vert \vec{p}\right\vert }\left(  z\right)  =\sum_{n=0}^{\infty}%
\frac{\left(  \frac{-\Omega_{\Lambda}}{18}z^{3}\right)  ^{n}\Gamma\left(
1-i\frac{\left\vert \vec{p}\right\vert }{3}\right)  }{n!\Gamma\left(
n+1-i\frac{\left\vert \vec{p}\right\vert }{3}\right)  }=~_{0}F_{1}\left(
1-i\frac{\left\vert \vec{p}\right\vert }{3},\frac{-\Omega_{\Lambda}z^{3}}%
{18}\right) \\
S_{\left\vert \vec{p}\right\vert }\left(  z\right)  \overset{z\rightarrow
\pm\infty}{\rightarrow}\frac{\Gamma\left(  1-i\frac{\left\vert \vec
{p}\right\vert }{3}\right)  e^{\frac{\pi\left\vert \vec{p}\right\vert }{6}}%
}{2\sqrt{\pi}(-1)^{-1/4}}\left(  \frac{\Omega_{\Lambda}z^{3}}{18}\right)
^{-\frac{1}{4}+i\frac{\left\vert \vec{p}\right\vert }{6}}\left[  e^{\frac
{i\pi}{4}+\frac{\pi\left\vert \vec{p}\right\vert }{6}-2i\sqrt{\frac
{\Omega_{\Lambda}z^{3}}{18}}}+e^{-\frac{i\pi}{4}-\frac{\pi\left\vert \vec
{p}\right\vert }{6}+2i\sqrt{\frac{\Omega_{\Lambda}z^{3}}{18}}}\right]  +\cdots
\end{array}
\label{case1}%
\end{equation}
The hypergeometric function $_{0}F_{1}\left(  a,W\right)  ,$ is an entire
function in the finite complex $W$ plane for any complex $a.$ The asymptotic
property of this $S_{\left\vert \vec{p}\right\vert }\left(  z\right)  $ was
used in order to fix the correct relative coefficients in Eqs.(\ref{BasA}%
,\ref{WaveUniverse},\ref{H+-}).

Other analytic solutions of interest in various limits of the $\Omega_{i}$
include the following
\begin{equation}%
\begin{array}
[c]{l}%
\text{Case-2:\ }\left(  \Omega_{\Lambda},\Omega_{K},p_{1},p_{2}\right)
\rightarrow0,\;V\left(  z\right)  \rightarrow\tilde{V}\left(  z\right)
=-\left(  \frac{p_{3}^{2}+1}{4z^{2}}+\frac{\Omega_{c}}{4z}\right)  ,\;\\
\mathcal{H}\Psi=\left(  \partial_{\phi}^{2}-\partial_{h}^{2}+\Omega
_{c}\right)  \Psi=0,\\
S_{\left\vert p_{3}\right\vert }\left(  z\right)  =\sum_{n=0}^{\infty}%
\frac{\left(  -\frac{\Omega_{c}}{4}z\right)  ^{n}\Gamma\left(  1-i\left\vert
p_{3}\right\vert \right)  }{n!\Gamma\left(  n+1-i\left\vert p_{3}\right\vert
\right)  }=~_{0}F_{1}\left(  1-i\left\vert p_{3}\right\vert ,\frac{-\Omega
_{c}~z}{4}\right)  .
\end{array}
\label{case2}%
\end{equation}
This is the case with $\hat{p}=\left(  0,0,\pm1\right)  $ that was studied in
\cite{barsAraya-rindler} (no anisotropy in this case). The WdWe (\ref{WdWfh})
for case-2 simplifies a great deal in the global basis $\left(  \phi,h\right)
$ as above. This \textquotedblleft massive\textquotedblright\ Klein-Gordon
equation in 1+1 dimensions is easily solved. The alternative bases, namely
Minkowski $\left(  \phi,h\right)  $ versus Rindler $\left(  z,\sigma\right)
$, are fully equivalent to each other in the classical theory. However, in the
quantum theory there is a non-trivial analyticity property in connecting
$\left(  \phi,h\right)  \leftrightarrow\left(  z,\sigma\right)  $ due to
branch points and branch cuts in $\Psi\left(  u,v\right)  .$ The unique
definition of $\Psi\left(  u,v\right)  $ introduces an infinite number of
sheets in the complex $u$ and $v$ planes. This is interpreted as a new
\textit{multiverse for which an extensive discussion }is given in
\cite{barsAraya-rindler}. The new multiverse is present in the general case
all $\Omega_{i}\neq0$.

A case with vanishing $\Omega_{\Lambda}\rightarrow0$, but all other parameters
non-vanishing, is
\begin{equation}%
\begin{array}
[c]{l}%
\text{Case 3:\ }\Omega_{\Lambda}\rightarrow0,\;V\left(  z\right)
\rightarrow\tilde{V}\left(  z\right)  =-\left(  \frac{\vec{p}^{2}+1}{4z^{2}%
}+\frac{\Omega_{c}}{4z}-\frac{\Omega_{K}}{4}\right)  ,\;\\
\mathcal{H}\Psi=\left(  \partial_{\phi}^{2}-\partial_{h}^{2}-\frac
{\partial_{1}^{2}+\partial_{2}^{2}}{\phi^{2}-h^{2}}-\Omega_{K}\left(  \phi
^{2}-h^{2}\right)  +\Omega_{c}\right)  \Psi=0.\\
S_{\left\vert \vec{p}\right\vert }\left(  z\right)  =\left\{
\begin{array}
[c]{l}%
=e^{-\frac{1}{2}\sqrt{\Omega_{K}}z}\sum_{n=0}^{\infty}\frac{\Gamma\left(
\frac{1-i\left\vert \vec{p}\right\vert }{2}-\frac{\Omega_{c}}{4\sqrt
{\Omega_{K}}}+n\right)  \Gamma\left(  1-i\left\vert \vec{p}\right\vert
\right)  }{\Gamma\left(  \frac{1-i\left\vert \vec{p}\right\vert }{2}%
-\frac{\Omega_{c}}{4\sqrt{\Omega_{K}}}\right)  \Gamma\left(  1-i\left\vert
\vec{p}\right\vert +n\right)  }\frac{\left(  \sqrt{\Omega_{K}}z\right)  ^{n}%
}{n!}\\
=e^{-\frac{1}{2}\sqrt{\Omega_{K}}z}~_{1}F_{1}\left[  (\frac{1-i\left\vert
\vec{p}\right\vert }{2}-\frac{\Omega_{c}}{4\sqrt{\Omega_{K}}}),\left(
1-i\left\vert \vec{p}\right\vert \right)  ,\sqrt{\Omega_{K}}z\right]
\end{array}
\right.  ,
\end{array}
\label{case3}%
\end{equation}
Here $_{1}F_{1}\left(  a,b,W\right)  $ is another hypergeometric function that
is entire in the complex $W$ plane. The asymptotic behavior of the potential
$\tilde{V}\left(  z\right)  $ is now dominated by the curvature constant
$\Omega_{K}.$

The classical solution analogous to Fig.7, namely $\left(  z\left(
\tau\right)  ,\vec{s}\left(  \tau\right)  \right)  $ turns out to be
completely periodic in this case, and gives the generic trajectory in Fig.8 as
compared to Fig.7. %

\begin{center}
\includegraphics[
height=1.6706in,
width=1.6577in
]%
{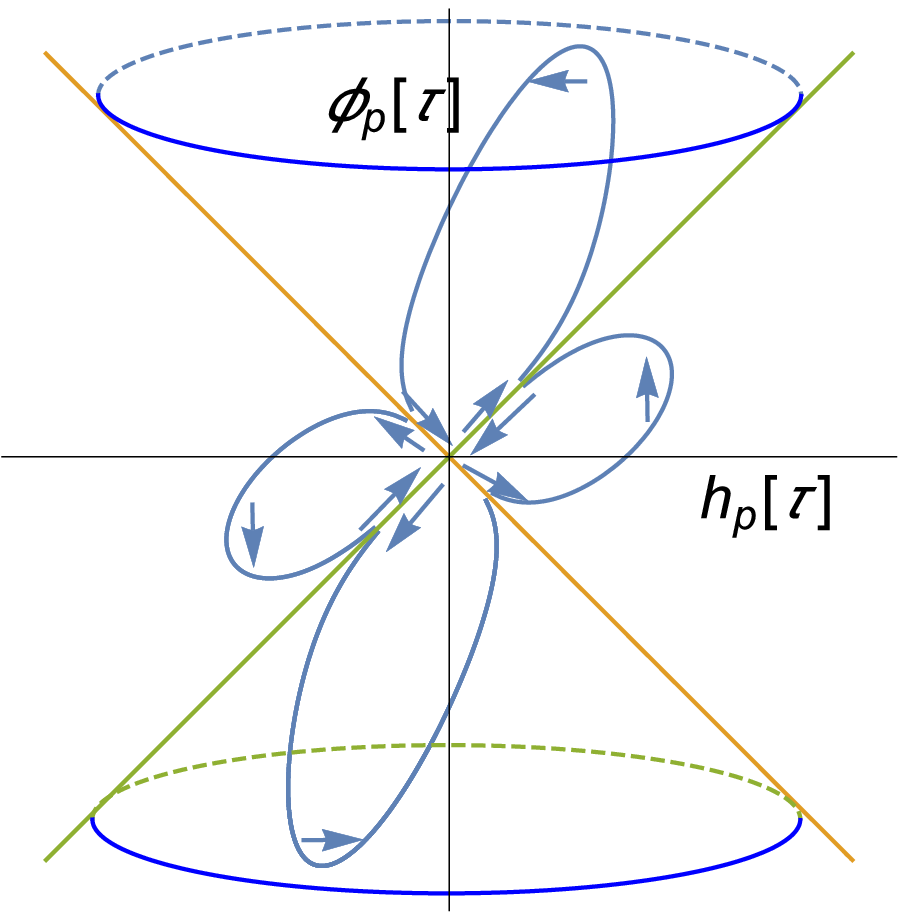}%
\\
{\protect\scriptsize Fig.8 - Periodic trajectory driven by curvature.}%
\end{center}
The analytic solution for this trajectory is given in terms of ordinary
periodic functions, and of course agrees with the $\Omega_{\Lambda}%
\rightarrow0$ limit of Eqs.(\ref{zLambda}-\ref{stau})%
\begin{equation}%
\begin{array}
[c]{l}%
z\left(  \tau\right)  =-\frac{2\vec{p}^{2}}{\sqrt{\Omega_{c}^{2}+4\Omega
_{K}\vec{p}^{2}}+\Omega_{c}}+\frac{\sqrt{\Omega_{c}^{2}+4\Omega_{K}\vec{p}%
^{2}}}{\Omega_{K}}\sin^{2}\left(  \sqrt{\Omega_{K}}\tau\right)  ,\;\\
\vec{s}\left(  \tau\right)  =\vec{s}_{0}+\frac{\vec{p}}{2\left\vert \vec
{p}\right\vert }\ln\left\vert \frac{\left(  \sqrt{\Omega_{c}^{2}+4\Omega
_{K}\vec{p}^{2}}+\Omega_{c}\right)  \tan\left(  \sqrt{\Omega_{K}}\tau\right)
-2\left\vert \vec{p}\right\vert \sqrt{\Omega_{K}}}{\left(  \sqrt{\Omega
_{c}^{2}+4\Omega_{K}\vec{p}^{2}}+\Omega_{c}\right)  \tan\left(  \sqrt
{\Omega_{K}}\tau\right)  +2\left\vert \vec{p}\right\vert \sqrt{\Omega_{K}}%
}\right\vert .
\end{array}
\label{zs-case3}%
\end{equation}

Another case is the specialized version of case-3 with vanishing anisotropy,
\begin{equation}%
\begin{array}
[c]{l}%
\text{Case 4:\ }\left(  \Omega_{\Lambda},p_{1},p_{2}\right)  \rightarrow
0,\;V\left(  z\right)  \rightarrow\tilde{V}\left(  z\right)  =-\left(
\frac{p_{3}^{2}+1}{4z^{2}}-\frac{\Omega_{K}}{4}\right)  ,\\
\left[  \partial_{\phi}^{2}-\partial_{h}^{2}-\Omega_{K}\left(  \phi^{2}%
-h^{2}\right)  +\Omega_{c}\right]  \Psi=0.
\end{array}
\label{case4}%
\end{equation}
Of course, $S_{\left\vert \vec{p}\right\vert }\left(  z\right)  $ is just the
$\left(  p_{1},p_{2}\right)  \rightarrow0$ limit of (\ref{case3}) and $\left(
z\left(  \tau\right)  ,\sigma\left(  \tau\right)  \right)  $ are just the
limits of (\ref{zs-case3}) when $p_{1}=p_{2}=s_{10}=s_{20}=0.$ However, just
like case-2, the WdWe has a nice interpretation in the global $\left(
\phi,h\right)  $ space in Fig.1: the WdWe (\ref{WdWfh}) reduces to the1+1
dimensional relativistic harmonic oscillator Hamiltonian that is constrained
to a single energy eigenvalue fixed to $-\Omega_{c}$, as indicated in
(\ref{case4}). The complete ghost-free unitary analysis of this equation was
first given in \cite{barsHO}, its application to cosmology was outlined in a
footnote in \cite{BCTsolutions}, and given again in detail in \cite{barsJames}
where it was noted $\Omega_{c}$ must be quantized in this setting, a path
integral quantization was also applied in \cite{Turok-Gillen} that is in
complete agreement with \cite{barsJames} but misses on the quantization of
$\Omega_{c}.$

Exactly solvable cases-(1-4) are of course only limits of the full case
discussed in this paper. These limits are helpful in better understanding the
structure and meaning of the full solution. But it must be noted that in some
of these limits, in particular those dominated by curvature (cases 3,4), and
those lacking anisotropy, the solution is qualitatively different.

Finally, I would like to present the wavefunction very close to the
\textquotedblleft beginning\textquotedblright, i.e. at the big bang. This is
not an input, but rather it is a unique prediction driven by the dynamics that
attracts to the pin hole in Fig.7 discussed earlier. $\Psi\left(  z,\vec
{s}\right)  $ close to the singularity is obtained by taking the
$z\rightarrow0$ limit of the general solution (\ref{WaveUniverse}). In this
expression there is no need to know the details of $S_{\left\vert \vec
{p}\right\vert }\left(  z\right)  $ because at $z=0$ it is exactly 1 in all
cases, including the full case with all non-zero parameters, $S_{\left\vert
\vec{p}\right\vert }\left(  0\right)  =1.$ The next to the leading terms,
\begin{equation}
S_{\left\vert \vec{p}\right\vert }\left(  z\right)  =1-\frac{\Omega_{c}%
}{4\left(  1-i\left\vert \vec{p}\right\vert \right)  }z+\frac{\left(
\Omega_{c}^{2}+4\Omega_{K}\left(  1-i\left\vert \vec{p}\right\vert \right)
\right)  }{32\left(  1-i\left\vert \vec{p}\right\vert \right)  \left(
2-i\left\vert \vec{p}\right\vert \right)  }z^{2}+O(z^{3}), \label{corrections}%
\end{equation}
are obtained from the series expansion in (\ref{case3}), while at order
$O\left(  z^{3}\right)  $ the parameter $\Omega_{\Lambda}$ also contributes as
seen from (\ref{case1}). I will concentrate only on the first term
$S_{\left\vert \vec{p}\right\vert }\left(  z\right)  \rightarrow1$ and
evaluate the integral $\Psi_{II}\left(  z,\vec{s}\right)  $ \textit{in the
proximity of the tip of the upper cone in Fig.7} (or 8)%
\begin{equation}
\Psi_{II}\left(  z,\vec{s}\right)  \overset{z\sim0}{=}\int_{-\infty}^{\infty
}d^{3}p\frac{Ae^{-\frac{p_{3}^{2}}{2\Omega_{\sigma}}-\frac{p_{1}^{2}+p_{2}%
^{2}}{2\Omega_{\alpha}}-i\vec{p}\cdot\vec{s}}}{\sqrt{\left(  2\pi\right)
^{3}2\left\vert \vec{p}\right\vert }}\left(  \sqrt{\left\vert z\right\vert
}^{-i\left\vert \vec{p}\right\vert }+\sqrt{\left\vert z\right\vert
}^{i\left\vert \vec{p}\right\vert }\frac{e^{-\frac{\pi\left\vert \vec
{p}\right\vert }{3}}\left(  \frac{\Omega_{\Lambda}}{18}\right)  ^{i\frac
{\left\vert \vec{p}\right\vert }{3}}\Gamma\left(  -i\frac{\left\vert \vec
{p}\right\vert }{3}\right)  }{\Gamma\left(  i\frac{\left\vert \vec
{p}\right\vert }{2}\right)  }\right)  .
\end{equation}
I will approximate $\Omega_{\sigma}=\Omega_{\alpha}\equiv\Omega_{\sigma
,\alpha}$ to have a rotationally symmetric integrand that is simpler to
evaluate. This is sufficient to get the general idea. Then the angular
integration over $\hat{p}$ yields $\int d^{2}\hat{p}e^{-i\vec{p}\cdot\vec{s}%
}=4\pi e^{-i\left\vert \vec{p}\right\vert \left\vert \vec{s}\right\vert },$
while the remaining radial integral is a function of only $\left(  z,s\right)
,$ with $s=\sqrt{\alpha_{2}^{2}+\alpha_{2}^{2}+\sigma^{2}}>0.$ The second term
in the parenthesis oscillates wildly as $z\rightarrow0,$ and it vanishes, as
expected in the discussion of Eq.(\ref{WaveUniv}) near $z=0$. However, the
first term has a stationary region when $z\rightarrow0$ and $s\rightarrow
\infty$ in tandem so that $e^{s}\sqrt{\left\vert z\right\vert }$ is finite.
This is the attractor mechanism at work, showing once again that anisotropy
and scalar degrees of freedom diverge while the scale factor vanishes. I
find,
\begin{equation}
\Psi_{II}\left(  z,s\right)  \overset{z\sim0}{=}\frac{4\pi A}{\sqrt{\left(
2\pi\right)  ^{3}}}\int_{0}^{\infty}\frac{p^{2}dp}{\sqrt{2p}}~e^{-\frac{p^{2}%
}{2\Omega_{\sigma,\alpha}}}\left(  e^{s}\sqrt{\left\vert z\right\vert
}\right)  ^{-ip}=\frac{A\pi^{2}\left(  -1\right)  ^{\frac{3}{4}}}%
{\sqrt{2\left(  2\pi\right)  ^{3}}\Omega_{\sigma,\alpha}^{7/2}}\phi\left(
x\right)  .
\end{equation}
The integral is performed exactly. It is written in terms of Bessel functions
$I_{\nu}\left(  -\frac{x^{2}}{4}\right)  $, and expressed as a function
$\phi\left(  x\right)  $ of a finite and positive $x$,%
\begin{equation}
\phi\left(  x\right)  \equiv\sqrt{x}e^{-\frac{x^{2}}{4}}\left[
\begin{array}
[c]{c}%
\left(  x^{2}-1\right)  I_{-\frac{1}{4}}\left(  -\frac{x^{2}}{4}\right)
+\left(  x^{2}-3\right)  I_{\frac{1}{4}}\left(  -\frac{x^{2}}{4}\right) \\
+x^{2}\left(  I_{\frac{3}{4}}\left(  -\frac{x^{2}}{4}\right)  +I_{\frac{5}{4}%
}\left(  -\frac{x^{2}}{4}\right)  \right)
\end{array}
\right]  ,\;x\equiv\frac{e^{\left\vert \vec{s}\right\vert }\sqrt{\left\vert
z\right\vert }}{\sqrt{\Omega_{\sigma,\alpha}}}.
\end{equation}
The $x$ parameter is proportional to some average of the $\left(  u_{\hat{p}%
},v_{\hat{p}}\right)  $ parameters in Fig.7. This $\phi\left(  x\right)  $ is
a complex function whose absolute value and phase are plotted in Fig.9. The
upper curve in Fig.9, $\left\vert \phi\left(  x\right)  \right\vert ,$
represents the probabilistic distribution of the wavefunction near the tip of
the cone in Fig.7. It shows that the probability is larger for smaller values
of $\left(  u_{\hat{p}},v_{\hat{p}}\right)  .$ This means larger probability
when close to the pin hole, consistent with the classical solution in Fig.7
that shows horizons are crossed \textit{classically only at the pin hole}.
Fig.9 conveys a fuzzy quantum version of the same result, confirming that
anisotropy and Higgs degrees of freedom must get larger $\left(  \left\vert
\vec{s}\right\vert \rightarrow\infty\right)  $ as the universe gets smaller
$\left(  z\rightarrow0\right)  $ so that probability continuously propagates
through the \textit{neighborhood of the pin hole}. This further softening of
singularities is due to quantum mechanics.%

\begin{center}
\includegraphics[
height=1.558in,
width=2.5589in
]%
{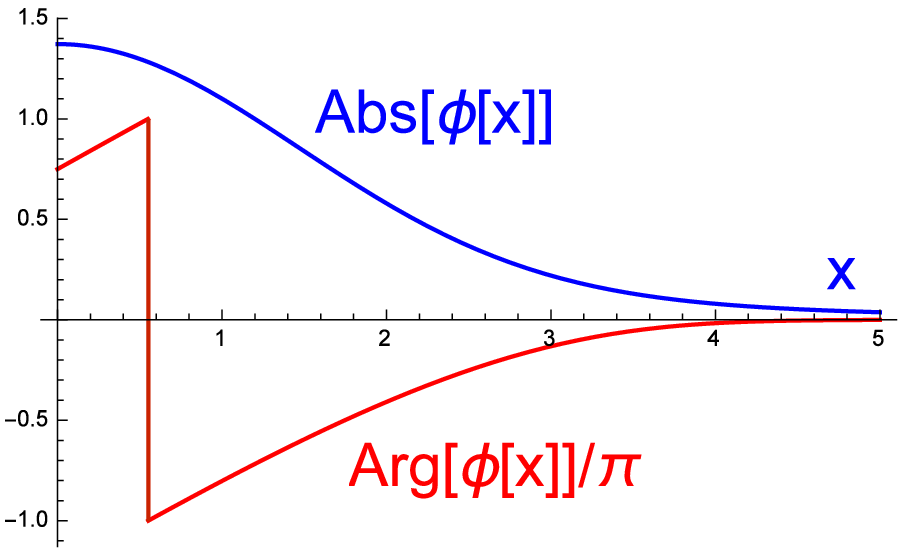}%
\\
{\protect\scriptsize Fig.9 - Wavefunction near pin hole.}%
\end{center}

Predicted corrections to this result near $z\simeq0$ follow directly from
(\ref{corrections}). The wavefunction for all $\left(  z,\vec{s}\right)  $
throughout the geodesically complete superspace is also determined in
Eq.(\ref{WaveUniverse}), and can be plotted in a similar way.

\section{Discussion and Outlook \label{outlook}}

In this section I will highlight the main results in this paper, comment on
open questions and problems, and compare to other discussions in recent
literature [\cite{neilEtAl1}-\cite{pirsaEtc}] that relate to the quantum
treatment of mini-superspace.

\subsection{Overview of results}

I defined an EFTC that is mathematically sufficiently well behaved at
gravitational singularities. Part of its quantitative definition includes the
suppression of higher curvature terms in the effective action by relying on
the softening effects of some underlying theory of quantum gravity (QG). The
remaining singular terms are mathematically controllable with a local scale
symmetry. Quantum mechanics makes it even softer, since passage through a
singularity occurs in a \textit{neighborhood of a point} rather than only at a
point, as demonstrated quantitatively in Fig.9 for the \textit{predicted
initial conditions of the wavefunction of the universe at the big bang}.

Advocating that higher curvature terms be banned in an effective field theory
for cosmological applications is a new point of view that is motivated by
notions of QG. Undoubtedly, this blurry point that seems reasonable at the
outset, needs more discussion, and I hope it will be a starting point for
future investigations and improvements. The temporary justification for this
EFTC is that it provides a practically working formalism to investigate
quantitatively singular spacetimes, including cosmology and black holes, and
be able to make predictions that were not available before. Furthermore, at
the fundamental level, this EFTC is grounded in the successful Standard Model
and General Relativity, with only a modest improvement to achieve geodesic
completeness of its spacetime through a local scale (Weyl) symmetry. Combined
with the softer classical and quantum mathematical properties, this provides a
physically strong basis for new progress whose results can be compared to
other attempts of QG when those can produce comparable computations.

This paper is focussed on quantizing the degrees of freedom of
mini-superspace, including scale, Higgs, anisotropy, dark matter and dark
energy, radiation and curvature, that are expected to play the main roles in
shaping the very early universe and its later development. One aim was
extracting from them the prediction of this EFTC for the wavefunction of the
universe. This was fulfilled in this paper, culminating in the prediction of
an explicit wavefunction of the universe that contains no parameters, has
dynamically produced unique initial values, and is continuous in a
geodesically complete universe that includes gravity as well as antigravity patches.

It is straightforward to compute the Feynman propagator from any point to any
other point in the geodesically complete superspace as follows%
\begin{equation}%
\begin{array}
[c]{l}%
G\left(  \phi^{\prime},h^{\prime},\alpha_{1,2}^{\prime};\phi,h,\alpha
_{1,2}\right)  =\langle\phi^{\prime},h^{\prime},\alpha_{1,2}^{\prime}|\frac
{i}{\mathcal{H}+i\varepsilon}|\phi,h,\alpha_{1,2}\rangle\\
\;\;=i\int d\lambda~\left(  \lambda+i\varepsilon\right)  ^{-1}\Psi_{\lambda
}\left(  \phi^{\prime},h^{\prime},\alpha_{1,2}^{\prime}\right)  \Psi_{\lambda
}^{\ast}\left(  \phi,h,\alpha_{1,2}\right)  .
\end{array}
\label{propagator}%
\end{equation}
where $\mathcal{H}\Psi_{\lambda}=\lambda\Psi_{\lambda}$ is explicitly given by
the WdWe differential operator in either (\ref{WdWfh}) or (\ref{WdWezs}) and
by replacing $\Omega_{c}$ by $\left(  \Omega_{c}-\lambda\right)  .$ A quick
(but not always the best version) solution for $\Psi_{\lambda}$ is,
$\Psi_{\lambda}=\Psi_{\lambda=0}\left(  \Omega_{c}-\lambda\right)  ,$ where
$\Psi_{\lambda=0},$ is the complete set of solutions of the WdWe already
obtained in the previous sections. The result of (\ref{propagator}) can be
expressed in either the global coordinates $\Psi_{\lambda}\left(
\phi,h,\alpha_{1,2}\right)  $ or the patchy coordinates $\Psi_{\lambda}\left(
z,\vec{s}\right)  $. In the latter case, the choice of $\Psi_{II}$ or
$\Psi_{I\&III}$ or $\Psi_{IV}$ depends on the location of the corresponding
\textquotedblleft points\textquotedblright\ $\left(  \phi^{\prime},h^{\prime
},\alpha_{1,2}^{\prime};\phi,h,\alpha_{1,2}\right)  $ in the geodesically
complete mini-superspace. Although these details are not fully carried out
here for all values of the parameters $\left(  \left\vert \vec{p}\right\vert
,\Omega_{\Lambda},\Omega_{c},\Omega_{K}\right)  $, the complete solution is
already available in the literature for some sub-cases. This is thanks to the
recognition emphasized in Eq.(\ref{rind}), that the glabal mini-superspace
$\left(  z,\sigma\right)  $ has the geometry of Rindler space geodesically
completed to 1+1 dimensional flat Minkowski space. Then some computations
become very simple. Namely, the following propagators,

\begin{description}
\item[(a)] Case-2 in Eq.(\ref{case2}) is equivalent to the massive
Klein-Gordon (KG) equation in 1+1 dimensions, so the associated propagator is
simply the massive KG propagator in the full space in Fig.1.
\begin{equation}
G\left(  \phi^{\prime},h^{\prime};\phi,h\right)  =i\int\frac{dp_{\phi}dp_{h}%
}{\left(  2\pi\right)  ^{2}}\frac{\exp\left(  -i\left(  \phi^{\prime}%
-\phi\right)  p_{\phi}+i\left(  h^{\prime}-h\right)  p_{h}\right)
}{\left(
-p_{\phi}^{2}+p_{h}^{2}+\Omega_{c}+i\varepsilon\right)  }.\label{prop-2}%
\end{equation}
This can be expressed in terms of the patchy coordinates $G\left(
z^{\prime },\sigma^{\prime};z,\sigma\right)  $ by the coordinate
transformation (\ref{coordTransf}). It is harder to compute the
propagator directly in the $\left(  z,\sigma\right)  $ basis using
the $z$-version of $\mathcal{H}$ in Eq.(\ref{case2}) including the
potential $\tilde{V}\left(  z\right)  $. However, with some labor
involving Bogoliubov transformations between Rindler waves and
Minkowski waves, given in \cite{barsAraya-rindler}, the propagator
can be brought to this form.

\item[(b)] Case-4 in Eq.(\ref{case4}), is equivalent to the relativistic
harmonic oscillator, whose unitary infinite dimensional quantum basis is given
in \cite{barsHO}. The assocociated propagator is worked out explicitly in
\cite{barsJames}. This can again be easily rewritten in terms of $\left(
z,\sigma\right)  ,$ and in that form the propagator is generalized to case-3
in the $\left(  z,\vec{s}\right)  $ basis. The case-3 propagator will appear
in a separate paper.
\end{description}

The computations in this paper were possible thanks to the control provided at
cosmological singularities by the underlying local scaling (Weyl) gauge
symmetry in this EFTC. The new computations at both classical and quantum
levels revealed surprizing dominant behavior of some degrees of freedom at the
very beginning of the universe. Specifically, this led to a theorem that
states: \textit{anisotropy and Higgs (or another scalar) degrees of freedom
must keep growing indefinitely as the scale factor of the universe keeps
getting smaller when crunch or bang type singularities are approached.} This
striking conclusion was derived in the quantum treatment of the wavefunction
through the steps of sections (\ref{QM1},\ref{QM2},\ref{QM3}). Specifically,
it is because of this behavior that the wavefunction manages to be continuous
in propagating through cosmological singularities that separate gravity and
antigravity patches. This quantum conclusion strengthens an earlier similar
result in the classical treatment \cite{crunchBang} of the relevant degrees of
freedom. With the inclusion of the cosmological constant that was missing in
\cite{crunchBang}, this paper presents a more complete unique classical
solution in section (\ref{attractor}) that displays a spectacular attractor
mechanism and passage through the singularity, as represented by the pin hole
in Fig.7. Furthermore, the computations also predicted the mathematically
unavoidable \textit{multiverse aspects of the wavefunction} (more thoroughly
discussed in \cite{barsAraya-rindler}); the multiverse continues to be under
study to understand its physical significance.

It should not go unnoticed that the Higgs in this EFTC has important
cosmological roles in shaping the very early universe. These include,
providing geodesic completeness, participating in the attractor mechanism and
continuity of the wavefunction from gravity to antigravity patches, and in the
avoidance of the mixmaster chaos (last part of section \ref{attractor}).

\subsection{Open problems}

There are open questions that deserve further investigation:

(\textrm{i}) Anisotropy is predicted to be huge at the beginning, then how
does it become miniscule in today's universe? Some would advocate inflation as
a possible mechanism, but inflation has not yet been considered as an added
feature to this EFTC, although such a modification of the EFTC may be
considered as an option. However, it is interesting that a very different and
rather natural mechanism has also emerged in this paper for how anisotropy can
evolve from huge at the bang to tiny today. The basic idea is the observation
enunciated just before Eqs.(\ref{Vapprox}-\ref{packet}) that motivated the
2-step strategy for taking into account approximately the effect of the
potentials $V_{K}\left(  \alpha_{1},\alpha_{2}\right)  $, $V\left(
\sigma,\varepsilon_{z}\right)  .$ Namely, in a time dependent Hamiltonian,
degrees of freedom that are subjected to attractive potentials, will quickly
descend to the ground state. In the case of anisotropy, the time (i.e. $z$)
dependent potential is $\left(  \phi^{2}-h^{2}\right)  V_{K}\left(  \alpha
_{1},\alpha_{2}\right)  =zV_{K}\left(  \alpha_{1},\alpha_{2}\right)  ,$ as
seen in (\ref{fAndVK},\ref{Sminifh}). A plot of $V_{K}$ \cite{misner} shows
that this is an infinite potential well, of the approximate shape of an upside
down infinite triangular pyramid, whose strength $\left(  \phi^{2}%
-h^{2}\right)  =z$ keeps growing as the universe expands. The progressively
stronger attractive potential will bind anisotropy more and more tightly in
its ground state, thus driving $\alpha_{1,2}\rightarrow0$. This seems like a
perfect natural mechanism to explain why the average\textit{ homogeneous
anisotropy} is so small in the later universe even though it is infinitely
large at the bang. It should be mentioned that in discussions of dynamics in
this potential \cite{BKL}\cite{inhomog}, it is claimed that not only average
anisotropy but also average inhomogeneity (if included in the equations in the
first place) would tend to get smaller as the universe expands. Renewed
vigorous investigations, on whether this scenario actually produces sufficient
suppression of anisotropy as well as inhomogeneities, and the extent to which
this supports the 2-step strategy applied in this paper, would be useful.

(ii) A similar investigation regarding the 2-step strategy with the Higgs
potential is in order. The reasoning was that at the electroweak phase
transition the Higgs should settle to the minimum of its potential,
$\left\vert h/\phi\right\vert \rightarrow w\sim10^{-17},$ and remain there for
the subsequent evolution. This seems reasonable in the gravity sectors where
$\left\vert h/\phi\right\vert <1.$ However during evolution in the antigravity
sector where $\left\vert h/\phi\right\vert >1,$ the Higgs potential $V\left(
\phi,h\right)  =\left(  \phi^{2}-h^{2}\right)  ^{2}f\left(  h/\phi\right)  $
given in (\ref{fAndVK}) is far from the minimum; then the huge $\Omega
_{\lambda}\sim10^{120}$ term can play an important role. The modification of
$\Psi$ can be assessed qualitatively as follows: The huge $\Omega_{\lambda}$
term creates a very strong potential that prevents $h$ from getting large
during antigravity; then the loops in the antigravity sectors in Fig.7 (and
correspondingly the probability amplitude $\left\vert \Psi_{I\&III}\right\vert
$ at large $\left\vert z\right\vert $) will be considerably smaller since
$\left\vert z\right\vert =h^{2}\left\vert 1-\left(  \phi/h\right)
^{2}\right\vert $ is prevented more strongly from growing in antigravity$.$
There may be other interesting effects during antigravity that are hard to
guess without an explicit computation. In addition, recall that any function
$f\left(  h/\phi\right)  $ is consistent with Weyl symmetry; even sticking
with the standard model form in (\ref{fAndVK}), the renormalized dimensionless
parameters $\left(  \Omega_{\lambda},w,\Omega_{\Lambda}\right)  $ as functions
of $\ln\left(  h/\phi\right)  ,$ are not known beyond lowest order in
perturbation theory. This remaining unknown in the effective theory can be a
source of speculation, including the possibility of a metastable Higgs
\cite{metastable} with additional dramatic consequences in our understanding
of cosmology, as discussed in \cite{BST-Higgs}. Further studies that address
such remaining questions related to the Higgs would be of interest.

(iii) The next goal is to include \textit{inhomogeneous} small fluctuations of
the metric and Higgs in the mini-superspace action (\ref{smini-general}) and
treat them as perturbations to the unique homogeneous background solution for
$\Psi_{II,I\&III,IV}$ given in section (\ref{QM3}). The results would
eventually be confronted with available data and phenomenology of the observed
properties of the cosmic microwave background (CMB). The background
wavefunction reported in this paper $\Psi_{II,I\&III,IV}\left(  z,\vec
{s}\right)  $ already has built in non-perturbative dominant parts of the
metric such as anisotropy, as well as matter such as the Higgs, as part of the
$\vec{s}$ dependence. Therefore the suggested expansion in small fluctuations
of the metric and Higgs is very different than previous attempts, either in
the usual classical approach, or the few quantum versions attempted in recent
literature. This is because the huge non-perturbative dynamical effects, such
as $\left(  \left\vert \vec{s}\right\vert \rightarrow\infty\text{ when
}z\rightarrow0\right)  ,$ were not known or taken into account in previous
computations. Because of this, I expect that the \textit{previous
computations, that as part of their setup, assumed only perturbative small
fluctuations, should have internal inconsistencies}.

\subsection{Comparing quantum approaches}

The last remark provides an introduction to a comparison of the current work
to other recent path integral approaches \cite{neilEtAl1}-\cite{pirsaEtc} that
have discussed the quantum mini-superspace. Because there is some confusing
debate still brewing in this topic, it would be useful to readers to clarify
where the present work stands relative to this controversy. A main message is
that the other approaches lack some of the important and essential features in
the current paper and there is room for improvement of the path integral
computations if these features can (I believe some difficulty) incorporated:

\begin{enumerate}
\item The authors in \cite{neilEtAl1}-\cite{pirsaEtc} use path integral
quantization as opposed to WdWe method to compute the wavefunction of the
universe or a related propagator. In principle all such methods should agree,
so different approaches are welcome. As in item $\left(  b\right)  $ above, I
find agreement for the \textit{propagator} in case-4 Eq.(\ref{case4}) in the
WdWe method \cite{barsJames} versus the counterpart in the path integral
method \cite{Turok-Gillen}. This is a good sign, but beyond this, so far,
there is little available in the other approaches to compare with the results
in the current paper. As a next easy comparison, I would suggest the
propagator in Eq.(\ref{prop-2}) for case-2 in (\ref{case2}), that is not
available yet in the path integral formalism.

\item This paper presents exact quantum solutions for the WdWe and its
propagators. By contrast, the path integral results are only semi-classical.
Sharp disagreements between competing groups, \cite{neilEtAl1}%
-\cite{neilEtAl4} versus \cite{neilEtAl1}-\cite{halliwellEtAl4}, doing path
integral computations remain unsettled. Part of the controversy is over the
fundamental correctness of Lorentzian versus Euclidean path integration in the
computation of the wavefunction for the universe. On that score, I side with
Lorentzian as a principle, but also the agreement of propagators in
\cite{barsJames} versus \cite{Turok-Gillen} noted in item $\left(  1\right)  $
lends support to Lorentzian. A second, more subtle technical part of the
controversy, involves which path is the correct integration path, to define
the quantum theory - this should be settled by comparing to the WdWe work in
this paper. A third part of the controversy is that one group claims to
compute a wavefunction while the other group insists on propagators. The
current paper based on the WdWe approach produces exact quantum results for
both quantities. Future semiclassical path integral results that may disagree
with the exact quantum results of the current paper would, in my opinion, be suspect.

\item The path integral teams have been working with a geodesically incomplete
mini-superspace that covers only region II in Figs.1 or 7. Signals of the
incompleteness arises in their computations; specifically, their parameter
$q>0,$ that is related directly to $q=a_{E}^{2}=\left\vert z\right\vert ,$
runs into contradictions with the mathematical properties of their equations
because imposing $q>0$ (half space of my $z$) at the quantum level is
problematic; but they sweep this problem under the rug. For the purpose of
comparing to their results, it is possible to narrow the results of the
present work to only region II, and therein there are fundamental differences
of principle. In particular initial conditions\textit{ at the big bang} is
really an input in their case (even though it is called \textquotedblleft no
boundary proposal\textquotedblright), but it is an output and a prediction in
the current paper as seen in Fig.9 and related equations. This difference is
connected to the attractor mechanism in Figs.6,7 that is completely lacking in
their approach because the drivers of this mechanism are absent in their
simplified model.

\item Most importantly, the path integral approaches do not include some of
the mini-superspace degrees of freedom. Specifically, anisotropy, scalar
field, conformal dust matter $\Omega_{c},$ are hugely dominant in the early
universe as compared to the cosmological constant $\Omega_{\Lambda}$. Yet, in
the models investigated in the recent path integral papers, the cosmological
constant is the main ingredient driving the evolution. Leaving out certain
terms in the action produces a much more manageable integral, but
unfortunately this misses the \textit{dominant non-perturbative effects of
anisotropy and Higgs} emphasized repeatedly from different classical and
quantum perspectives in the current paper.

\item In the Lorentzian path integral approach, inconsistencies concerning
small \textit{inhomogeneous metric fluctuations} were discovered
\cite{neilEtAl1}-\cite{neilEtAl4}. As reported, computation shows that the
fluctuations come out larger than the \textit{homogeneous} background;
however, this is contrary to the setup of the computation in which
\textit{linearized fluctuations} were assumed to be smaller than the
background to begin with. As noted at the end of item (iii) above, this is to
be expected since, as this paper demonstrated, there are very large
non-perturbative effects in the homogeneous metric and scalar field, namely
anisotropy and Higgs, that should be part of the background. Inhomogeneous
metric and scalar fluctuations, on top on this non-perturbative background,
would be expected to remain consistently small and overcome this problem.
\end{enumerate}

Meaningful comparison between the results of the current paper and the path
integral approach will be possible, for both the wavefunction and the
propagator, when the listed differences in the approaches are ironed out.
These include the choice of models and degrees of freedom they contain,
inclusion of potentials and/or implementation of the 2-step strategy for
approximating their effects, semiclassical versus exact quantum computation,
and the inclusion of non-perturbative effects in the homogeneous background
solution. Given the encouraging agreement for the propagator in one of the
simplest cases (case-4 in Eq.(\ref{case4})) as reported in \cite{barsJames}
and \cite{Turok-Gillen}, I expect full agreement when the computations of
various groups focus on the same system and the same physical quantities.

\subsection{Towards an ultraviolet completion of the EFTC}

The geodesically complete EFTC promoted in this paper is capable of providing
detailed quantitative description of passage through cosmological
singularities at both the classical and quantum levels. This kind of
prediction is also possible for black holes by using the same EFTC as
suggested in \cite{barsArayaJames}. In this way it is demonstrated that events
in the spacetime on the other side of singularities (such as far past boundary
conditions) affect the properties of the physics in the spacetime past the
singularities. So, the geodesically complete spacetime must be taken into
account for cosmology. When this is done, as in this paper, initial conditions
at the big bang are \textit{predicted} not guessed. This in itself is
remarkable about this EFTC because other approaches in cosmology (including
stringy approaches) have not been able to provide comparable detail.

As emphasized earlier in this paper there are three crucial ingredients in
this EFTC: $\left(  i\right)  $ a Weyl symmetry and associated geodesic
completeness, $\left(  ii\right)  $ a ban on higher derivatives at high
energies and $\left(  iii\right)  $ close connection to the Standard Model at
low energies, including the Higgs field. I now address the question of
\textquotedblleft how could these ingredients be compatible with an
ultraviolet complete approach, such as string theory\textquotedblright?

The ban on higher derivatives is in fact attributed to a softening provided by
quantum gravity, such as string theory. This is expected just on the basis
that the description of the physics in the strongly interacting regime is
given in terms of stringy configurations involving string fields (including
stringy modes) as compared to point-like fields in the low energy
approximation. Past experience with string theory shows that perturbative
string amplitudes expanded in powers of $\alpha^{\prime}$ (string tension of
Planck scale) are reproduced in the low energy effective theory by including
higher derivative terms (such as higher curvatures) that are multiplied by
powers of $\alpha^{\prime}.$ But these terms are valid only at small momenta
(small derivatives) or small energies $E$ when $\alpha^{\prime}E^{2}\ll1.$ The
string amplitudes at high energies $\alpha^{\prime}E^{2}\gtrsim1$ cannot be
reproduced by using the higher derivative terms of the low energy theory.
Furthermore, the stringy description at high energy does not involve higher
curvatures, but instead it involves stringy modes that provide a much softer
behavior of the theory even in a strongly interacting regime. Therefore, it is
completely wrong to include higher derivative terms in the low energy theory
if the purpose is to describe a phenomenon such as the transitions through a
singularity. This is the justification for banning higher derivatives in the EFTC.

The EFTC is of course not a substitute for an ultraviolet complete theory. At
best, the EFTC is expected to correctly describe the physics up to some
fraction of the Planck energy and possibly be inaccurate at higher energies.
Nevertheless it is gratifying that the EFTC in this paper does provide a
mathematically self-consistent answer to questions at the scale of Planck
energies, including passage through singularities. It presently stands as the
only tool that provides quantitative answers to questions at the Planck scale.
Until a more reliable tool becomes available, I believe this is at least an
answer to think about. To be certain of the physical correctness or
inaccuracies of this EFTC description at the Planck scale one must construct
and then analyze an appropriate string theory and then compare answers.

Unfortunately string cosmology is not an easy task. Past attempts have
encountered a number of difficulties, including those described in sec.3.4 of
\cite{burgess} and references therein. Part of the problem is that most
attempts rely on the perturbative setup of string theory for strings that
propagate on a cosmological background. However, near a singularity stringy
interactions become strong so that a perturbative stringy approach cannot
work. For instance near null cosmological singularities (which allow for
detailed analysis), strings are known to become highly excited (nontrivial
oscillator modes) suggesting that backreaction is important. Likewise in
discussions of cosmological singularities and gauge/gravity duality, it can be
argued in certain cases (with spacelike singularities) that the dual gauge
theory (which might have been naively hoped to lead to a controlled weakly
coupled description) also breaks down, implying that continuing past the bulk
Big-Crunch singularity is unclear (the AdS Kasner singularities of \cite{Das}
have been revisited in \cite{horowitz} and subsequent work). If a smooth gauge
theory description exists allowing continuation past the singularity (e.g. as
certain null singularities suggest), it would amount to the bulk gravitational
description necessarily being strongly coupled. These and related
investigations remain inconclusive themselves but point to some difficulties
of the perturbative setup in string cosmology and showing that the
perturbative setup to string cosmology is useless. A more useful, but quite
difficult, approach could be string field theory, where non-perturbative
solutions in terms of string fields in an appropriate cosmological background
are possible, thus possibly providing a better non-perturbative tool.

In any case, I believe that, in both perturbative and non-perturbative string
theory, to capture the correct physics one must use geodesically complete
backgrounds that include all patches of a complete spacetime on both sides of
singularities. That this is essential has been demonstrated in this paper in
the context of the EFTC. However, the notion of a geodesically complete space
is totally missing in all previous attempts in string cosmology. Connected to
the same fact, the notion of a stringy background that is also Weyl symmetric
in target space has also been missing in overall string theory because string
theory has a fundamental length $\alpha^{\prime}$ (the string torsion related
to the Newton constant $G_{N}$). On the other hand, since the Weyl symmetric
EFTC does exist as in this paper, in which $G_{N}$ is generated by spontaneous
breakdown, and geodesic completeness is built in, one should wonder which
string background could yield it as a low energy approximation?

The previous paragraph poses a challenge for all quantum gravity (QG)
attempts, not only string theory. The low energy EFTC in the present paper is
geodesically complete, and the Weyl symmetry is crucial. On the other hand all
known attempts for QG, including string theory, have a dimensionful parameter
that is equivalent to the gravitational constant $G_{N}$, so they are not Weyl
symmetric in target spacetime, and do not have an effective gravitational
function that could change sign so that geodesic completeness of the
backgrounds is built in. Clearly, such inherently geodesically incomplete QG
theories could not generate the EFTC suggested in this paper. However, it is
possible to improve string theory with Weyl symmetry on target space to make
it consistent with the properties of this EFTC. This is possible by replacing
the string tension in string theory to be a background field that can change
sign, as shown in \cite{nontrivialString}. How this can be incorporated in the
BRST operator in string field theory has also been discussed briefly in
\cite{barsJames}.

The Weyl-improved string theory is of course difficult to analyze, but at
least it has the right properties to be the ultraviolet completion of the EFTC
discussed in this paper. Future work may reach a stage that provides stringy
results to be compared to those obtained in this paper, thus showing the level
of success or shortcomings of the EFTC.

\begin{acknowledgments}
I thank the Perimeter Institute for its generous support and hospitality
during my sabbatical. Research at the Perimeter Institute is supported by the
Government of Canada and the Province of Ontario through the Ministry of
Research and Innovation.
\end{acknowledgments}

\end{document}